\documentclass[titlepage,11pt]{article}

\usepackage{amsthm}
\usepackage{amsfonts}
\usepackage{amsmath}
\usepackage{amssymb}
\usepackage{graphicx}
\usepackage{lineno}
\usepackage{setspace} %\onehalfspacing %Makes 1.5-spacing.
\usepackage{latexsym}
\usepackage{amsfonts}
\usepackage{amsmath}
\usepackage{verbatim}
\usepackage{hyperref}
\usepackage{enumerate}

\newcommand\blackslug{\hbox{\hskip 1pt \vrule width 4pt height 8pt depth 1.5pt
        \hskip 1pt}}
\newcommand\bbox{\hfill \quad \blackslug \bigbreak}
\def\d{\hbox{-}}

\def\l{,\ldots,}

\title{Edge-colouring seven-regular planar graphs}
\author{Maria Chudnovsky\thanks{Supported by NSF grants DMS-1001091 and IIS-1117631.}\\
Columbia University, New York, NY 10027
\\
\\
Katherine Edwards\thanks{Supported by an NSERC PGS-D3 Fellowship and a Gordon Wu Fellowship.}\\
Princeton University, Princeton, NJ 08544
\\
\\
Ken-ichi Kawarabayashi\thanks{Partially supported by the Mitsubishi
Foundation.}\\
National Institute of Informatics and JST ERATO Kawarabayashi Project, Japan
\\
\\
Paul Seymour\thanks{Supported by ONR grant N00014-10-1-0680 and NSF grant DMS-0901075.}\\
Princeton University, Princeton, NJ 08544}

\date{August 17, 2012; revised \today}

\newtheorem{thm}{}[section]

\newcommand{\Proof}{\noindent{\bf Proof.}\ \ }

\begin{document}
\maketitle
\begin{abstract}
A conjecture due to the fourth author states that every $d$-regular 
planar multigraph 
can be $d$-edge-coloured, provided that for every odd set $X$
of vertices, there are at least $d$ edges between $X$ and its complement.
For $d = 3$ this is the four-colour theorem, and the conjecture has been 
proved for all $d\le 8$, by various authors. In particular, two of us proved it when $d=7$; and then three of us proved it when $d=8$. The methods used for the latter
give a proof in the $d=7$ case that is simpler than the original, and we present it here.
\end{abstract}

\section{Introduction}
Let $G$ be a graph. (Graphs in this paper are finite, and may have loops or parallel edges.)
If $X\subseteq V(G)$, $\delta_G(X) = \delta(X)$ denotes the set of all
edges of $G$ with an end in $X$ and an end in $V(G)\setminus X$.
We say that $G$ is {\em oddly $d$-edge-connected} if $|\delta(X)|\ge d$
for all odd subsets $X$ of $V(G)$. 
The following
conjecture~\cite{seymour} was proposed by the fourth author in about 1973.

\begin{thm}\label{mainconj}
{\bf Conjecture.} If $G$ is a $d$-regular planar graph, then $G$ is 
$d$-edge-colourable if and only if $G$ is oddly $d$-edge-connected.
\end{thm}

The ``only if'' part is true, and some special cases of the ``if'' part of this conjecture have been proved. 
\begin{itemize}
\item For $d = 3$ 
it is the four-colour theorem, and was
proved by Appel and Haken~\cite{appelhaken1,appelhaken2,rsst}; 
\item for $d = 4,5$ it was proved by Guenin~\cite{guenin}; 
\item for $d = 6$ it was proved by Dvorak, Kawarabayashi
and Kral~\cite{dvorak}; 
\item for $d = 7$ it was proved by the second and third authors and appears in the Master's thesis~\cite{katie} of the former; 
\item for $d = 8$ it was proved by three of us~\cite{ces12}.
\end{itemize}

The methods of \cite{ces12} can be adapted to yield a proof of the result for $d=7$, that is shorter and simpler than that of \cite{katie}. 
Since in any case the original proof appears only in a thesis, we give the new one here. 
Thus, we show
\begin{thm}\label{mainthm}
Every $7$-regular oddly $7$-edge-connected planar graph is $7$-edge-colourable.
\end{thm}

All these proofs (for $d>3$), including ours, proceed by induction on $d$. 
Thus we need to assume the truth of the result for $d = 6$. 
Some things that are proved in ~\cite{ces12} are true for any $d$, and we sometimes cite results from that paper.

\section{An unavoidable list of reducible configurations.}

% The graph we wish to edge-colour has parallel edges, but it is more convenient to work with the underlying simple graph.
% If $H$ is $d$-regular and oddly $d$-edge-connected, then $H$ has no loops, because
% for every vertex $v$, $v$ has degree $d$, and yet $|\delta_H(v)|\ge d$. (We write $\delta(v)$ for $\delta(\{v\})$.)
% Thus to recover $H$ from the underlying simple graph $G$ say, we just need to know the number $m(e)$ of parallel edges
% of $H$ that correspond to each edge $e$ of $G$. 
Let us say a {\em $d$-target} is a pair $(G,m)$
with the following properties
(where for $F\subseteq E(G)$, $m(F)$ denotes $\sum_{e\in F} m(e)$):
\begin{itemize}
\item $G$ is a simple graph drawn in the plane;
\item $m(e)\ge 0$ is an integer for each edge $e$;
\item $m(\delta(v)) = d$ for every vertex $v$; and
\item $m(\delta(X))\ge d$ for every odd subset $X\subseteq V(G)$.
\end{itemize}

In this language, \ref{mainconj} says that for every $d$-target $(G,m)$, there is a list of $d$ perfect matchings of $G$
such that every edge $e$ of $G$ is in exactly $m(e)$ of them. 
(The elements of a list need not be distinct.) If there is such a list we call it a {\em $d$-edge-colouring}, and say that $(G,m)$ is {\em
$d$-edge-colourable}.
For an edge $e\in E(G)$, we call $m(e)$ the {\em multiplicity} of $e$.
If $X\subseteq V(G)$, $G|X$ denotes the subgraph of $G$ induced on $X$.
We need the following theorem from \cite{ces12}:

\begin{thm}\label{counterex}
Let $(G,m)$ be a $d$-target, that is not $d$-edge-colourable, but such that every $d$-target with fewer vertices is $d$-edge-colourable. Then 
\begin{itemize}
\item $|V(G)|\ge 6$;
\item for every $X\subseteq V(G)$ with $|X|$ odd, if $|X|,|V(G)\setminus X|\ne 1$ then $m(\delta(X))\ge d+2$; and
\item $G$ is three-connected, and $m(e)\le d-2$ for every edge $e$.
\end{itemize}
\end{thm}

\bigskip

A {\em triangle} is a region of $G$ incident with exactly three edges. If a triangle is incident with vertices $u,v,w$, for convenience we refer to it as
$uvw$, and in the same way an edge with ends $u,v$ is called $uv$.
Two edges are {\em disjoint} if they are distinct and no vertex is an end of both of them, and otherwise they {\em meet}.
Let $r$ be a region of $G$, and let $e\in E(G)$ be incident with $r$; let $r'$ be the other region incident with $e$. We say that $e$
is {\em $i$-heavy} (for $r$), where $i\ge 2$, if either
$m(e)\ge i$ or $r'$ is a triangle $uvw$ where $e = uv$ and
$$m(uv)+\min(m(uw),m(vw))\ge i.$$
We say $e$ is a {\em door} for $r$ if $m(e) = 1$ and there is an edge $f$ incident with $r'$ and disjoint from $e$ with $m(f) = 1$.
We say that $r$ is {\em big} if there are at least four doors for $r$, and {\em small} otherwise. A {\em square} is a region with length four.

Since $G$ is drawn in the plane and is two-connected, every region $r$ has boundary some cycle which we denote by $C_r$.
In what follows we will be studying cases in which certain configurations of regions are present in $G$.
We will give a list of regions the closure of the union of which is a disc. For convenience, for an edge $e$
in the boundary of this disc, we call the region outside the disc incident with $e$ the ``second region'' for $e$; and we write $m^+(e) = m(e)$ if the second region
is big, and $m^+(e) = m(e)+1$ if the second region is small. This notation thus depends not just on $(G,m)$ but on what regions we have specified, so it is
imprecise, and when there is a danger of ambiguity we will specify it more clearly.
If $r$ is a triangle, incident with edges $e,f,g$, we define its {\em multiplicity} $m(r) = m(e) + m(f) + m(g)$. We also write $m^+(r) = m^+(e) + m^+(f) + m^+(g)$.
A region $r$ is {\em tough} if $r$ is a triangle and $m^+(r)\geq 7$.

Let us say a $7$-target $(G,m)$ is {\em prime} if 
\begin{itemize}
\item $m(e)>0$ for every edge $e$;
\item $|V(G)|\ge 6$;
\item $m(\delta(X))\ge 9$ for every $X\subseteq V(G)$ with $|X|$ odd and $|X|,|V(G)\setminus X|\ne 1$;
\item $G$ is three-connected, and $m(e)\le 6$ for every edge $e$;
\end{itemize}
and in addition $(G,m)$ contains none of of the following:
\begin{enumerate}[{\bf Conf(1):}]
{\item\label{c1} A triangle $uvw$, where $u$ has degree three and its third neighbour $x$ satisfies $$m(ux)<m(uw)+m(vw).$$}
{\item\label{c2} Two triangles $uvw,uwx$ with $m(uv) + m(uw) + m(vw)+ m(ux)\ge 7$.}
{\item\label{c3} A square $uvwx$ where $m(uv)+m(vw)+m(ux)\ge 7$.}
{\item\label{c4} Two triangles $uvw,uwx$ where $m^+(uv)+m(uw)+m^+(wx) \ge 6$.}
{\item\label{c5} A square $uvwx$ where $m^+(uv)+m^+(wx)\ge 6$.}
{\item\label{c6} A triangle $uvw$ with $m^+(uv)+m^+(uw) = 6$ and either $m(uv)\geq 3$ or $m(uv)=m(vw)=m(uw)=2$ or $u$ has degree at least four.}
{\item \label{c7} A region $r$ of length at least four, an edge $e$ of $C_r$ with $m^+(e)=4$ where every edge of $C_r$ disjoint from $e$ is $2$-heavy and not incident with a triangle with multiplicity three, and such that at most three edges disjoint from $e$ are not $3$-heavy.}
{\item \label{c8} A region $r$ with an edge $e$ of $C_r$ with $m^+(e)=m(e)+1=4$ and an edge $f$ disjoint from $e$ with $m^+(f)=m(f)+1=2$, where every edge of $C_r\setminus\{f\}$ disjoint from $e$ is $3$-heavy with multiplicity at least two.}
{\item \label{c9} A region $r$ of length at least four and an edge $e$ of $C_r$ such that $m(e)=4$ and there is no door disjoint from $e$. Further for every edge $f$ of $C_r$ consecutive with $e$ with multiplicity at least two, there is no door disjoint from $f$.}
{\item \label{c10} A region $r$ of length four, five or six and an edge $e$ of $C_r$ such that $m(e)=4$ and such that $m^+(f)\geq 2$ for every edge $f$ of $C_r$ disjoint from $e$.}
{\item\label{c11} A region $r$ and an edge $e$ of $C_r$, such that $m(e) = 5$ and at most five edges of $C_r$ disjoint from $e$ are doors for $r$, or $m^+(e)=m(e)+1=5$ and at most four edges of $C_r$ disjoint from $e$ are doors for $r$.}
{\item\label{c12} A region $r$, an edge $uv$ of $C_r$, and a triangle $uvw$ such that $m(uv) + m(vw)=5$ and at most five edges of $C_r$ disjoint from $v$ are doors for $r$.}
{\item \label{c13} A square $xuvy$ and a tough triangle $uvz$, where $m(uv)+m^+(xy)\geq 4$ and $m(xy)\geq 2$.}
{\item \label{c14} A region $r$ of length five, an edge $f_{0}\in E(C_{r})$ with $m^{+}(e_{0})\geq 2$ and $m^{+}(e)\geq 4$ for each edge $e\in E(C_{r})$ disjoint from $f_{0}$.}
{\item \label{c15} A region $r$ of length five, a $3$-heavy edge $f_{0}\in E(C_{r})$ with $m(e_{0})\geq 2$ and $m^{+}(e)\geq 3$ for each edge $e\in E(C_{r})$ disjoint from $f_{0}$.}
{\item \label{c16} A region $r$ of length six where five edges of $C_r$ are $3$-heavy with multiplicity at least two.}
\end{enumerate}

We will prove that $7$-target is prime (Theorem~\ref{unav}). 
To deduce \ref{mainthm}, we will show that if there is a counterexample, then some counterexample is prime; but
for this purpose, just choosing a counterexample with the minimum number of vertices is not enough, and we need a more delicate minimization.
If $(G,m)$ is a $d$-target, its {\em score sequence} is the $(d+1)$-tuple $(n_0,n_1 \l n_d)$ where $n_i$ is the number of edges
$e$ of $G$ with $m(e) = i$. If $(G,m)$ and $(G',m')$ are $d$-targets, with score sequences $(n_0 \l n_d)$ and $(n_0' \l n_d')$ respectively,
we say that $(G',m')$ is {\em smaller} than $(G,m)$ if either
\begin{itemize}
\item $|V(G')|< |V(G)|$, or
\item $|V(G')|= |V(G)|$ and there exists $i$ with $1\le i\le d$ such that $n_i'>n_i$, and $n_j' = n_j$ for all $j$ with $i<j\le d$, or
\item $|V(G')|= |V(G)|$, and $n_j' = n_j$ for all $j$ with $0<j\le d$, and $n_0'<n_0$.
\end{itemize}
% (The anomalous treatment of $n_0$ is just a device to allow $d$-targets to have edges with $m(e) = 0$, while minimum $d$-counterexamples have none.)
If some $d$-target is not $d$-edge-colourable, then we can choose a $d$-target $(G,m)$ with the following properties:
\begin{itemize}
\item $(G,m)$ is not $d$-edge-colourable
\item every smaller $d$-target is $d$-edge-colourable.
\end{itemize}
Let us call such a pair $(G,m)$ a {\em minimum $d$-counterexample}. 
To prove \ref{mainthm}, we prove two things:
\begin{itemize}
\item No $7$-target is prime (theorem~\ref{unav}), and
\item Every minimum $7$-counterexample is prime (theorem~\ref{reduc}).
\end{itemize}
It will follow that there is no minimum $7$-counterexample, and so the theorem is true.

\section{Discharging and unavoidability}

In this section we prove the following, with a discharging argument.

\begin{thm}\label{unav}
No $7$-target is prime.
\end{thm}

The proof is broken into several steps, through this section.
Let $(G,m)$ be a $7$-target, where $G$ is three-connected. For every region $r$, we define 
$$\alpha(r) = 14 - 7|E(C_r)| +  2\sum_{e\in E(C_r)}m(e).$$
We observe first:

\begin{thm}\label{totalsum}
The sum of $\alpha(r)$ over all regions $r$ is positive.
\end{thm}
\Proof
Since $(G,m)$ is a $7$-target, $m(\delta(v)) = 7$ for each vertex $v$, and, summing over all $v$, we deduce that
$2m(E(G)) = 7|V(G)|$. By Euler's formula, the number of regions $R$ of $G$ satisfies $|V(G)| - |E(G)| + R = 2$,
and so $4m(E(G)) - 14|E(G)| + 14R = 28$.
But $2m(E(G))$ is the sum over all regions $r$, of  $\sum_{e\in E(C_r)}m(e)$, and $14R-14|E(G)|$ is the sum over all regions $r$ of $14-7|E(C_r)|$.
It follows that the sum of $\alpha(r)$ over all regions $r$ equals $28$. This proves \ref{totalsum}.~\bbox

% We normally wish to pass one unit of charge from every small region to every big region with which it shares an edge; except that in some circumstances, sending one unit is not enough, and we send $2$. 
For every edge $e$ of $G$, define $\beta_e(s)$ for each region $s$ as follows.
Let $r,r'$ be the two regions incident with $e$. 

\begin{itemize}
\item If $s\ne r,r'$ then $\beta_e(s) = 0$.
\item If $r,r'$ are both big or both tough or both small and not tough, then  $\beta_e(r), \beta_e(r') = 0$.
\item[{\bf [$\beta$0]:}] If $r'$  is tough and $r$ is small and not tough then $\beta_e(r) = -\beta_e(r') = 1$.
\end{itemize}
Henceforth we assume that $r$ is big and $r'$ is small; let 
$f,g$ be the edges of $C_{r'}\setminus e$ that share an end with $e$.

\begin{enumerate}[{\bf [$\beta$1]:}]
  \item If $e$ is a door for $r$ (and hence $m(e) = 1$) then $\beta_e(r) = \beta_e(r') = 0$.
  % \item If $m(e)=1$ and $r'$ is a triangle with $m(f)=1$ and $m(f)\geq 3$ then $\beta_e(r) = -\beta_e(r') = \tfrac{3}{2}$.
  \item If $r'$ is a triangle with $m(r')\geq 5$ then $\beta_e(r) = -\beta_e(r') = 2$.
  \item Otherwise $\beta_e(r) = -\beta_e(r') = 1$.
\end{enumerate}

% (Think of $\beta_e$ as passing some amount of charge between the two regions incident with $e$.)
For each region $r$, define $\beta(r)$ to be the sum of $\beta_e(r)$ over all edges $e$. 
We see that the sum of $\beta(r)$ over all regions $r$ is zero. 

Let $\alpha, \beta$ be as above. Then the sum over all regions $r$ of $\alpha(r)+\beta(r)$ is positive,
and so there is a region $r$ with $\alpha(r)+\beta(r)>0$. Let us examine the possibilities for such a region.  
There now begins a long case analysis, and to save writing we just say ``by Conf(7)'' instead of ``since $(G,m)$ does not contain Conf(7)'', and so on.

\begin{thm}\label{bigovercharge}
If $r$ is a big region and $\alpha(r)+\beta(r)>0$, then $(G,m)$ is not prime.
\end{thm}
\Proof
Suppose that $(G,m)$ is prime.
Let $C = C_r$. Suppose $\alpha(r)+\beta(r)>0$; that is,
$$\sum_{e\in E(C)}(7-2m(e) - \beta_e(r)) < 14.$$
For $e\in E(C)$, define $\phi(e) = 2m(e) + \beta_e(r)$, and let us say $e$ is {\em major} if $\phi(e)>7$.
If $e$ is major, then since $\beta_e(r)\le 3$, it follows that $m(e)\ge 3$ and that $e$ is $4$-heavy. 
If $m(e) = 3$ and $e$ is major, then by Conf(\ref{c1}) the edges consecutive with $e$ on $C$ have multiplicity at most two. It follows that no two major edges are consecutive, since $G$ has minimum degree at least three. Further when $e$ is major, $\beta_e(r)$ is an integer from the $\beta$-rules, and therefore $\phi(e) \ge 8$. 

Let $D$ be the set of doors for $C$.
Let 

\begin{itemize}
\item $\xi = 2$ if there are consecutive edges $e,f$ in $C$ such that $\phi(e) >9$ and $f$ is a door for $r$,
\item $\xi = 3$ if not, but there are consecutive edges $e,f$ in $C$ such that $\phi(e) =9$ and $f$ is a door for $r$,
\item $\xi = 4$ otherwise.
\end{itemize}
(1) {\em Let $e,f,g$ be the edges of a path of $C$, in order, where $e,g$ are major. Then}
$$(7-\phi(e))+2(7-\phi(f))+(7-\phi(g)) \ge 2\xi|\{f\}\cap D|.$$
Let $r_1,r_2,r_3$ be the regions different from $r$ incident with $e,f,g$ respectively.
Now $m(e)\le 5$ since $G$ has minimum degree three, and if $m(e) = 5$ then $r_1$ is big, by Conf(\ref{c11}), and so $\beta_e(r) = 0$. If $m(e)=4$ then $\beta_e(r)\leq 2$; and so in any case, $\phi(e)\le 10$. Similarly $\phi(g)\le 10$.  Also, $\phi(e), \phi(g)\ge 8$ since $e,g$ are major. Thus $\phi(e)+\phi(g)\in \{16,17,18,19,20\}$.

Since $f$ is consecutive with a major edge, $m(f) \leq 2$.
% Also, $r_2$ is not a triangle with an edge of multiplicity at least $3$ so rule {\em $\beta 2$} does not apply. 
Further if $m(f)=2$ then $r_2$ is not a triangle with multiplicity at least $5$ by Conf(\ref{c1}) so rule {\em $\beta 2$} does not apply.
Therefore it follows from the $\beta$-rules that $\phi(f)\leq 5$ and if $m(f)=1$ then $\phi(f) \leq 4$.

First, suppose that one of $\phi(e),\phi(g)\ge 10$, say $\phi(e) = 10$. In this case we must show that $2\phi(f) \le 18- \phi(g)-2\xi|\{f\}\cap D|$. It is enough to show that $2\phi(f) \le 8-2\xi|\{f\}\cap D|$. Now $m(e)\geq 4$ and $e$ is $5$-heavy by the $\beta$-rules, and so $m(f)=1$, since $G$ is three-connected and by Conf(\ref{c1}). If $f$ is a door then $\phi(f) = 2$ by rule {\em $\beta 1$} and $\xi=2$ so $2\phi(f) \le 8-2\xi|\{f\}\cap D|$.
If $f$ is not a door then since $\phi(f) \leq 4$, it follows that $2\phi(f) \le 8-2\xi|\{f\}\cap D|$. So we may assume $\phi(e),\phi(g) \leq 9$.
%tight for xi=2

Next, suppose that one of $\phi(e),\phi(g) = 9$, say $\phi(e) = 9$. By the $\beta$-rules, we have $m^+(e) = m(e)+1=5$. We must show that $2\phi(f) \le 19 -\phi(g)-2\xi|\{f\}\cap D|$; it is enough to show $2\phi(f) \le 10 -2\xi|\{f\}\cap D|$. Since $\phi(f)\leq 5$ we may assume $f$ is a door. Thus $\phi(f)=2$ and $\xi\leq 3$, so $4 = 2\phi(f) \leq 19-\phi(g)-2\xi|\{f\}\cap D|$. We may therefore assume that $\phi(e) + \phi(g) = 16$.
%tight for xi=3

So, suppose $\phi(e) + \phi(g) = 16$ and so $\phi(e)=\phi(g) = 8$. Now $\xi\leq 4$ and we must show that $2\phi(f) \leq 12-2\xi|\{f\}\cap D|$. Again, if $f$ is not a door then $2\phi(f) \leq 12$ as required. If $f$ is a door then $2\phi(f) = 4 \leq 12 - 2\xi|\{f\}\cap D|$.
%not tight
This proves (1).
\\
\\
(2) {\em Let $e,f$ be consecutive edges of $C$, where $e$ is major. Then }
$$(7-\phi(e))+2(7-\phi(f)) \ge 2\xi|\{f\}\cap D|.$$
We have $\phi(e)\in \{8,9,10\}$. Suppose first that $\phi(e) = 10$. We must show that $2\phi(f)\le 11-2\xi|\{f\}\cap D|$; but $m(f)=1$ by Conf(\ref{c1}) since $e$ is $5$-heavy. Since $\phi(f)\leq 4$ we may assume $f$ is a door. Thus $\phi(f)=2$ and $\xi\leq 2$, as needed.
%By the $\beta$-rules $\beta_e(r)\leq \tfrac 32$, and so $\phi(f)\le 3.5$ as needed. %Thus $2\phi(f) \leq 11-2\xi|\{f\}\cap D|$.
%not tight

Next, suppose that $\phi(e) \leq 9$; it is enough to show that $2\phi(f)\le 12-2\xi|\{f\}\cap D|$. Now $e$ is $4$-heavy and $m(f)\leq 2$ so $\phi(f) \leq 6$ by the $\beta$-rules. We have $\xi \leq 4$. Since $\phi(f)\leq 6$, we may assume $f$ is a door. If $f$ is a door, then $2\phi(f) = 4 \leq 12-2\xi|\{f\}\cap D|$. This proves (2).
%tight

\bigskip

For $i = 0,1,2$, let $E_i$ be the set of edges $f\in E(C)$ such that $f$ is not major, and $f$ meets exactly $i$ major edges in $C$.
By (1), for each $f\in E_2$ we have 
$$\frac{1}{2}(7-\phi(e))+(7-\phi(f))+\frac{1}{2}(7-\phi(g))  \ge \xi|\{f\}\cap D|$$ 
where $e,g$ are the major edges meeting $f$. 
By (2), for each $f\in E_1$ we have
$$\frac{1}{2}(7-\phi(e))+(7-\phi(f)) \ge \xi|\{f\}\cap D|$$
where $e$ is the major edge consecutive with $f$.
Finally, for each $f\in E_0$ we have
$$7-\phi(f)\ge  \xi|\{f\}\cap D|$$
since $\phi(f)\le 7$, and $\phi(f) = 2$ if $f\in D$. Summing these inequalities over all $f\in E_0\cup E_1\cup E_2$, we deduce that
$\sum_{e\in E(C)}(7-\phi(e))\ge \xi|D|$.
Consequently 
$$14> \sum_{e\in E(C)}(7-2m(e) - \beta_e(r)) \ge \xi |D|.$$ 
% But $|D|\ge 4$ since $r$ is big, and so $\xi = 2$ and $|D|\le 6$, a contradiction by Conf(\ref{c11}) and Conf(\ref{c12}).
But $|D|\ge 4$ since $r$ is big, and so $\xi \leq 3$. If $\xi=3$, then $|D|=4$, contrary to Conf(\ref{c11}). So $\xi=2$ and $|D|\le 6$. But then $C_r$ has a $5$-heavy edge with multiplicity at least four that is consecutive with a door and has at most five doors disjoint from it, contrary to Conf(\ref{c11}) and Conf(\ref{c12}).
%Used |D|\ge 4$
This proves \ref{bigovercharge}.~\bbox

\begin{thm}\label{triovercharge1}
If $r$ is a triangle that is not tough, and $\alpha(r)+\beta(r)>0$, then $(G,m)$ is not prime.
\end{thm}
\Proof
Suppose $(G,m)$ is prime, and let $r = uvw$. Now $\alpha(r) = 2(m(uv)+m(vw)+m(uw))-7$,  so 
$$ 2(m(uv)+m(vw)+m(uw))+\beta(r) >7.$$
Let $r_1,r_2,r_3$ be the regions different from $r$ incident with $uv,vw,uw$ respectively.
Since $r$ is not tough, $m^+(r)\leq 6$, and so $m(r)\leq 6$ as well.

Suppose first that $r$ has multiplicity six and hence $\beta(r) > -5$. Then $r_1,r_2,r_3$ are all big. Suppose $m(uv)=4$. Then rule {\em $\beta 2$} applies to give $\beta(r)=-6$, a contradiction. Thus $r$ has at least two edges with multiplicity at least two. Rules {\em $\beta 2$} and {\em $\beta 3$} apply giving $\beta(r)\leq -5$, a contradiction.
% Suppose $m(uv)=3$. Then rules {\em $\beta 7$} and {\em $\beta 9$} apply, giving $\beta(r)=-5$, a contradiction. Thus we may assume $m(uv)=m(vw)=m(uw)=2$. But in this case rule {\em $\beta 5$} applies and $\beta(r)=-6$, a contradiction. 

Suppose $r$ has multiplicity five and so $\beta(r) > -3$. Then at least two of $r_1,r_2,r_3$ are big, say $r_2$ and $r_3$, and so $\beta_{vw}(r) + \beta_{uw}(r) \leq -2$. Consequently $\beta_{uv}(r)> -1$ so we may assume that $r_1$ is a tough triangle $uvx$. By Conf(\ref{c2}), $m(ux)=m(vx)=1$. Since $uvx$ is tough, $m(uv)\geq 2$. Suppose $m(uv)=3$. Then by Conf(\ref{c4}), $m^+(ux)=m^+(vx)=1$, contradicting the fact that $uvx$ is tough.
So $m(uv)=2$, $m(uvx)=4$ and we may assume $m(vw)=2$. But by Conf(\ref{c4}), $m^+(ux)=1$, contradicting the fact that $uvx$ is tough. 

Suppose $r$ has multiplicity four. Then $\beta(r)> -1$. Since $m^+(r)\leq 6$ we may assume that $r_1$ is big, so $\beta_{uv}(r) = -1$. Now if $r_2$ is tough then $\beta_{vw}(r)=1$, and otherwise $\beta_{vw}(r)\leq 0$. Thus by symmetry we may assume $r_2$ is a tough triangle $vwx$ and $r_3$ is small. 
Suppose that $m(uv)=2$. 
By Conf(\ref{c4}), $m^+(vx) + m(vw) + m(uw) + 1 \leq 5$. Also by Conf(\ref{c4}), $m(uv)+m(vw) + m^+(wx) \leq 5$. Since $m(uv) + m(vw) + m(uw) = 4$ it follows that $m^+(vx) + m(vw) + m^+(wx) \leq 5$, contradicting the fact that $vwx$ is tough.

Therefore we may assume that $r$ has multiplicity three. Now $\beta(r) >1$. By the rules, if $r_1$ is tough then $\beta_{uv}(r) =1$. If $r_1$ is big then $\beta_{uv}(r)= -1$. Otherwise $\beta_{uv}(r) =0$. By symmetry, it follows that $r_1,r_2,r_3$ are all small and we may assume that $r_1,r_2$ are tough triangles $uvx$ and $vwy$. It follows from Conf(\ref{c4}) that $m^+(vx), m^+(ux) \leq 2$. This contradicts the fact that $uvx$ is tough.
This proves \ref{triovercharge1}.~\bbox

\begin{thm}\label{triovercharge2}
If $r$ is a tough triangle with $\alpha(r)+\beta(r)>0$, then $(G,m)$ is not prime.
\end{thm}
\Proof
Suppose $(G,m)$ is prime, and let $r = uvw$.
Now $\alpha(r) = 2(m(uv)+m(vw)+m(uw))-7$,  so 
$$ 2(m(uv)+m(vw)+m(uw))+\beta(r)>7.$$
Let $r_1,r_2,r_3$ be the regions different from $r$ incident with $uv,vw,uw$ respectively. Since $r$ is small and tough, observe from the rules that $\beta_e(r) \leq 0$ for $e=uv,vw,uw$. 

\bigskip

Let $X =\{u,v,w\}$. Since $(G,m)$ is prime, it follows that $|V(G)\setminus X|\ge 3$, and so $m(\delta(X))\ge 9$.
But
$$m(\delta(X)) = m(\delta(u)) +m(\delta(v))+ m(\delta(w)) - 2 m(uv)-2m(uw)-2m(vw),$$
and so $9\le 7+7+7 - 2 m(uv)-2m(uw)-2m(vw)$, that is,
$r$ has multiplicity at most six. Since $m^+(r)\geq 7$, $r$ has multiplicity at least four.

We claim that no two tough triangles share an edge. For suppose $uvw$ and $uvx$ are tough triangles.  By Conf(\ref{c4}), $m^+(vx) + m(uv) + m^+(uw) \leq 5$. Also by Conf(\ref{c4}) $m^+(vw) + m(uv) + m^+(ux) \leq 5$. Since $m^+(vw) + m^+(uw) + m(uv) \geq 6$, $m^+(vx) + m^+(ux) + m(uv) \leq 4$, contradicting the fact that $r_1$ is tough.

Suppose first that $r$ has multiplicity six and so $\beta(r) > -5$. By Conf(\ref{c2}), none of $r_1,r_2,r_3$ is a triangle. 
If $m(uv)=4$ then by Conf(\ref{c6}), $r_1,r_2,r_3$ are all big, contradicting the fact that $r$ is tough. 
If $m(uv)=3$, assume without loss of generality that $m(vw)=2$. Then by Conf(\ref{c6}), $r_1$ and $r_2$ are big, and rule {\em $\beta 2$} applies, contradicting that $\beta(r)> -5$. By symmetry we may therefore assume $m(uv)=m(vw)=m(uw)=2$. By Conf(\ref{c6}) we can assume $r_1$, $r_2$ are big and rule {\em $\beta 2$} applies again. This contradicts that $\beta(r) > -5$.

Consequently $r$ has multiplicity at most five. Then none of $r_1,r_2,r_3$ is tough and so $\beta(r) \leq =-3$, contradicting that $2(m(uv)+m(vw)+m(uw))+\beta(r)>7.$
This proves \ref{triovercharge2}.~\bbox

\begin{thm}\label{smallovercharge}

If $r$ is a small region with length at least four and with $\alpha(r)+\beta(r)>0$, then $(G,m)$ is not prime.
\end{thm}

\Proof
Suppose that $(G,m)$ is prime. Let $C = C_r$.
Since $\alpha(r) = 14 - 7|E(C)| +  2\sum_{e\in E(C)}m(e)$, it follows that 
$$14-7|E(C)| + 2\sum_{e\in E(C)}m(e) + \sum_{e\in E(C)}\beta_e(r)>0,$$
that is,
$$\sum_{e\in E(C)}(2m(e)+\beta_e(r)-7) > -14.$$
For each $e\in E(C)$, let 
$$\phi(e) = 2m(e)+\beta_e(r),$$ 
\\
\\
(1) {\em For every $e\in E(C)$, $\phi(e)\in \{1,2,3,4,5,6,7\} $.}
\\
\\
Since $r$ is not a triangle, $\beta_e(r)\in \{-1,0,1\}$. It follows from Conf(\ref{c11}) that $m(e)\leq 4$. Further, if $m(e)=4$ then $m^+(e)=4$ and $\beta_e(r)=-1$. This proves (1).

\bigskip

For each integer $i$, let $E_i$ be the set of edges of $C$ such that $\phi(e)=i$. From (1) $E(C)$ is the union of $E_1,E_2,E_3,E_4,E_5,E_6,E_7$.
\\
\\
Let $e$ be an edge of $C$ and denote by $r'$ its second region. We now make a series of observations that are easily checked from the $\beta$-rules and the fact that $2m(e)-1 \leq \phi(e) \leq 2m(e)+1$, as well as Conf(\ref{c6}) which implies that if $m(e)=3$ then $r'$ is not tough. %It will be useful to keep these in mind for the remainder of the proof.
\\
\\
(2) {\em $e\in E_1$ if and only if $m(e)=m^+(e)=1$ and $e$ is not a door for $r'$.}
\\
\\
(3) {\em $e\in E_2$ if and only if $m(e)=1$ and either
    \begin{itemize} 
      \item $m^+(e)=1$ and $e$ is a door for $r'$, or 
      \item $m^+(e)=2$ and $r'$ is not a tough triangle.
    \end{itemize}}
\noindent(4) {\em $e \in E_3$ if and only if either
    \begin{itemize} 
      \item $m(e)=1$ and $r'$ is a tough triangle, or 
      \item $m(e)=m^+(e)=2$.
    \end{itemize}}
\noindent(5) {\em  $e \in E_4$ if and only if $m(e)=2$, $m^+(e)=3$ and $r'$ is not a tough triangle.}
\\
\\
(6) {\em $e \in E_5$ if and only if either
    \begin{itemize}
      \item $m(e)=2$ and $r'$ is a tough triangle, or
      \item $m(e)=m^+(e)=3$.
    \end{itemize}}
\noindent(7) {\em $e \in E_6$ if and only if $m(e)=3$ and $m^+(e)=4$.}
\\
\\
(8) {\em $e \in E_7$ if and only if $m(e)=4$ and $m^+(e)=4$.}

\bigskip
\noindent(9) {\em No edge in $E_7$ is consecutive with an edge in $E_6\cup E_7$.}
\\
\\
Suppose that edges $e,f\in E(C)$ share an end $v$, and $e\in E_7$. Since $v$ has degree at least three it follows that $m(e) + m(f) \leq 6$ so $f\notin E_6\cup E_7$. This proves (9). 
\\
\\
(10) {\em Let $e$, $f$, $g$ be consecutive edges of $C$. If $e,g \in E_7$ then $f\in E_1\cup E_2 \cup E_3 \cup E_4$.}
\\
\\
For by (2), $f\notin E_6$. Suppose $f\in E_5$. Since $m(e)=m(g)=4$ and $G$ has minimum degree three, by (6) $m(f)=2$ and the second region for $f$ is a tough triangle $r'$ with $m(r')=4$. But $m^+(e)=m^+(g)=4$, so $r'$ is incident with two big regions; thus $m^+(r')=5$, contradicting the fact that $r'$ is tough. This proves (10).

\bigskip
For $1\le i\le 7$, let $n_i = |E_i|$. Let $k=|E(C)|$.
\\
\\
(11) {\em $5 n_1 +4n_2 +3n_3 + 2n_4 + n_5 + k  -n_7  \le 13$.}
\\
\\
Since
$$\sum_{e\in E(C)}(\phi(e)-7) > -14,$$
we have
$6 n_1 + 5 n_2 + 4 n_3 + 3 n_4 + 2 n_5 + n_6 \le 13$, that is, 

$$5 n_1 +4n_2 +3n_3 + 2n_4 + n_5 + k  -n_7  \le 13,$$
since $n_1+n_2+n_3+n_4+n_5 + n_6 + n_7 = k$, proving (11). 
\\
\\
(12) {\em $4n_1 + 3n_2 + 2n_3 + n_4 + k \leq 12$ and $n_1+n_2\le 2$.}
\\
\\
By (9) we have $n_1 + n_2 + n_3 + n_4 +n_5 \geq n_7$. Suppose $n_1 + n_2 + n_3 + n_4 +n_5 = n_7$. By Conf(\ref{c7}), the edges of $C$ cannot all be in $E_6$, so $n_7>0$. Then $k$ is even and every second edge of $C$ is in $E_7$, so by (10), $n_5=n_6=0$ so $n_1 + n_2 + n_3 + n_4 = \tfrac k2$ and $n_7=\tfrac k2$. By (11) $3n_1+2n_2+n_3 + \tfrac 32 k \leq 13$. Therefore, we either have $n_1+n_2\leq 1$, or $k=4$, or $n_1+n_2=2$ and $k=6$. But by Conf(\ref{c9}), every edge in $E_7$ is disjoint from an edge in $E_1\cup E_2$, a contradiction. This proves that $n_1 + n_2 + n_3 + n_4 +n_5 \geq n_7+1$. The first inequality follows from (11) and the second from the fact that $k\geq 4$. This proves (12).

\bigskip

\noindent{\bf Case 1:} $n_1+n_2=2$.

\bigskip
Suppose $k+n_1\geq 6$. By (12), $n_3=n_4=0$. By Conf(\ref{c9}), every edge in $E_7$ is disjoint from an edge in $E_1\cup E_2$, and therefore, by (9), is consecutive with an edge in $E_5$. Further, by (10) no edge in $E_5$ meets two edges in $E_7$, and so $n_5\geq n_7$, contradicting (11). This proves that $k+n_1 \leq 5$.

Suppose $k=5$. Then $n_2=2$, and so by (12), $n_3=0$ and $n_4\leq 1$. Also $n_4+ n_5+n_6+n_7= 3$. By (11), $n_7\geq 2n_4 + n_5$. Suppose $n_6=3$, then by (7), $C$ has three edges of multiplicity three, each of whose second region is small. Further, by (3) if the edges in $E_2$ are consecutive, they are both incident with small regions. This contradicts Conf(\ref{c14}). Therefore $n_6\leq 2$, and so $n_7\geq 1$. By Conf(\ref{c10}) one of the edges in $E_2$ must be incident with a big region and by (3), it must be a door for that region. Since $n_3=0$, it follows that the two edges in $E_2$ are disjoint. It follows that $n_7=1$. By (11), $n_4=0$ and $n_6\geq 1$. Let $e\in E_6$. Then $e$ must be consecutive with both edges in $E_2$, for it is not consecutive with the edge in $E_7$. But then $e$ is disjoint only from edges in $E_7\cup E_5$, contrary to Conf(\ref{c7}). 

Suppose that $k=4$. Then $n_1\leq 1$. By Conf(\ref{c10}) and (3), $n_1\geq n_7$. Therefore by (11), $3n_3 + 2n_4 + n_5 \leq 1$, and so $n_3=n_4=0$ and $n_5\leq 1$. Since $n_5+n_6+n_7=2$, and edges in $E_5, E_6, E_7$ have multiplicity at least two, three, four, respectively, Conf(\ref{c3}) implies $n_7=0$ and $n_6\leq 1$. Hence $n_5=n_6=1$. From (11) it follows that $n_1=0$. By Conf(\ref{c5}) the edge disjoint from the edge in $E_6$ must have multiplicity one and be incident with a big region. By (3) this edge must be in $E_1$, a contradiction.
This proves that Case 1 does not apply.

\bigskip

\noindent{\bf Case 2:} $n_1+n_2=1$.

\bigskip
Let $e_0 \in E_1\cup E_2$. %It follows from (2) and (3) that every edge of $C\setminus\{e_0\}$ is $2$-heavy. 
We claim that neither edge consecutive with $e_0$ is in $E_6\cup E_7$. For let $e_1$ be an edge consecutive with $e_0$ on $C$ and suppose $e_1 \in E_6\cup E_7$; then by (7), $m^+(e_1)=4$. But all edges disjoint from $e_1$ on $C$ are not in $E_1\cup E_2$ and therefore are $2$-heavy and their second regions are not triangles with multiplicity three. Therefore Conf(\ref{c7}) implies that at least four edges disjoint from $e_0$ are not $3$-heavy and hence $n_3+n_4\geq 4$ and that $k\geq 7$, contradicting (11). This proves that all edges in $E_6\cup E_7$ are disjoint from $e_0$, and so $n_3+n_4+n_5 \geq 2$. We consider two cases:
\bigskip

\noindent{\bf Subcase 2.1:} $n_7\geq 1$.
\\
Let $f\in E_7$. By Conf(\ref{c9}), if an edge $e_1$ meets both $e_0$ and $f$ then $m(e_1)=1$ and so $e_1\in E_3$. By (10) an edge meeting two edges in $E_7$ is in $E_3\cup E_4$. Summing over the edges meeting $E_7\cup\{e_0\}$ it follows that $2n_3 + 2n_4 + n_5 \geq 2(n_7+1)$.
From (11) we deduce $5n_1+4n_2+n_3 + n_7 +k\leq 11$; thus $k+n_1+n_3+n_7\leq 7$. By Conf(\ref{c10}), $m^+(e_0)=1$, so by (3), either $e_0\in E_1$ or there is an edge of multiplicity one disjoint from $e_0$. Since $n_1+n_2=1$, such an edge would be in $E_3$; it follows that $n_1+n_3\geq 1$. We deduce that $k\leq 5$.
If $k=5$ then by Conf(\ref{c9}) the edge meeting $e_0$ and $f$ is in $E_3$, and so $n_1+n_3\geq 2$, a contradiction. 

Thus $k=4$. Then by Conf(\ref{c10}) and (3), $e_0\in E_1$. By Conf(\ref{c3}) the two edges consecutive with $e_0$ are in $E_3$. But then $k+n_1+n_3+n_7 = 8$, a contradiction.
\\
\\
\noindent{\bf Subcase 2.2:} $n_7=0$.
\\
Let $e_0,\dots,e_{k-1}$ denote the edges of $C$ listed in consecutive order.
Since $n_3+n_4+n_5 \geq 2$, (11) implies $k\leq 7$.

Suppose $k=7$. Then the inequality in (11) is tight, and we have $n_2=1$, $n_5=2$ and $n_6=4$. Consequently $n_1=n_3=n_4=0$. Then $e_1, e_6\in E_5$, and so by (6) and (7) are $3$-heavy with multiplicity at least two, and $e_2,e_3,e_4,e_5 \in E_6$. This is a contradiction by Conf(\ref{c8}).

Suppose $k=6$. We know $e_1,e_5\notin E_6$. By (11),  $n_1 + 3n_3 + 2n_4 + n_5 \leq 3$, but $n_3+n_4+n_5\geq 2$ so $n_3=0$ and consequently $n_4+n_5+n_6=5$. Also $n_1+2n_4+n_5 \leq 3$. In particular $n_4\leq 1$. 
Suppose $n_4=1$, then $n_6=3$ and $n_5=1$ and $e_2, e_3, e_4 \in E_6$. It follows from Conf(\ref{c8}) that $m^+(e_0)=1$, and so $n_1=1$, contradicting that $n_1+2n_4+n_5 \leq 3$. 
Thus $n_4=0$. It follows that $n_5+n_6=5$. This contradicts Conf(\ref{c16}).

Next suppose $k=5$. We know $e_1,e_4\notin E_6$. By (11), $n_1 + 3n_3 + 2n_4 + n_5 \leq 4$. 
Suppose $2n_3+n_4\geq 2$. Then $n_1 + n_3+n_4+n_5 \leq 2$, and so $n_2 + n_6 \geq 3$. Since $n_6\leq 2$ we may assume $e_2,e_3\in E_6$ and $e_0\in E_2$,
contrary to Conf(\ref{c14}). 
It follows that $2n_3+n_4\leq 1$. Consequently $n_3=0$ and $n_5+n_6\geq 3$. Thus we may assume that $m^+(e_1), m^+(e_2), m^+(e_3), m^+(e_4) \geq 3$, and $e_1$ is $3$-heavy. This contradicts Conf(\ref{c15}).

%A region $r$ of length five where the edges of $C_r$ are $e_0,\dots,\e_4$ (listed in consecutive order), where $m^+(e_1), m^+(e_2), m^+(e_3), m^+(e_4) \geq 3$, and at least two of $e_1,e_2,e_3,e_4$ are $3$-heavy.

Finally, suppose $k=4$. By (11), $n_1 + 3n_3 + 2n_4 + n_5 \leq 5$. By Conf(\ref{c5}), at least one of $m^+(e_1), m^+(e_3) \leq 2$, so we may assume $e_1\in E_3$ and so $n_3=1$.  Since $m^+(e_1)=2$, Conf(\ref{c8}) implies $e_3\notin E_6$, and so $e_3 \in E_5$.
Suppose $e_0\in E_1$. Then $2n_4 + n_5 \le 1$, and so $n_4=0$ and $n_5\leq 1$. Since $e_2\notin E_5$, $e_2 \in E_6$. Since $m(e_2)=3$ by (7), it follows from Conf(\ref{c3}) that $m(e_1)=1$, $m(e_3)=2$ and from (4) and (6) that $e_1,e_3$ are incident with tough triangles $v_1v_2x$ and $v_3v_0y$. 
This contradicts Conf(\ref{c13}). 
% By Conf(\ref{c13}), $m^+(v_1x)=1$. Since $v_1v_2x$ is tough, $m^+(v_2x)\geq 4$, and so $m(v_2x)\geq 3$. This contradicts Conf(\ref{c1}).

Thus $e_0\in E_2$ and so $m^+(e_0)=2$. By Conf(\ref{c8}), $e_2 \notin E_6$. Hence $e_2\in E_4\cup E_5$. Since $2n_4+n_5 \leq 2$ and $e_3\in E_5$, it follows that $e_2 \in E_5$.
% Suppose $m(e_1)=1$, and so $e_1$ is incident with a tough triangle $v_1v_2x$. By Conf(\ref{c13}) (taking $v_1x$ as $v_0v_1$ in the configuration), $m^+(v_1x)=1$, and since $v_1v_2x$ is tough we have $m^+(v_2x)\geq 4$. But this contradicts Conf(\ref{c13}) (taking $v_2x$ as $v_0v_1$ in the configuration).
By Conf(\ref{c13}), the second region for $e_1$ is not a tough triangle, and so $m(e_1)=2$. 
Since $m(e_2),m(e_3) \geq 2$, Conf(\ref{c3}) tells us $m(e_3)=2$ and the second region for $e_3$ is a tough triangle $v_0v_3x$. 
But this contradicts Conf(\ref{c13}).
We conclude that Case 2 does not apply.
\bigskip

\noindent{\bf Case 3:} $n_1+n_2=0$.

\bigskip
In this case, $C$ has no doors, so by Conf(\ref{c9}) $n_7=0$. Suppose that $n_6\geq 1$ and let $e\in E_{6}$. Then by Conf(\ref{c7}), there are at least four edges disjoint from $e$ that are not $3$-heavy. Therefore $n_3+n_4 \geq 4$ and $k \geq 7$, contradicting (11). It follows that $n_{1} =n_{2}= n_{6} = n_{7}=0$, and so $n_3+n_4+n_5=k$. By (11), $3n_3 + 2n_4 + n_5 + k \leq 13$, and $k\leq 6$. Further, $3n_3 + 2n_4 + 2n_5 + k \leq 13 + n_5$, and so $n_5-n_3 \geq 3k-13$.

Suppose first that $k=6$; then $n_5\geq 5$, so by (6) $C$ has five $3$-heavy edges, each with multiplicity two or three, contrary to Conf(\ref{c16}). 
Suppose $k=5$; then $3n_3 + 2n_4 + n_5\leq 8$, and so, since $n_3+n_4+n_5=5$, $n_3\leq 1$. Also $n_5\geq 1$, and if $n_3=1$ then $n_4\leq 1$.
Consequently we may assume there is an ordering $e_0 \l e_4$ of $E(C)$, where $e_0\in E_5$ and $e_2,e_3\in E_4\cup E_5$, contrary to Conf(\ref{c15}).

Finally, suppose $k=4$; then $3n_3 + 2n_4 + n_5 \leq 9$. Since, by (5) and (6), every edge $f\in E_4\cup E_5$ has $m^+(f)\geq 3$, Conf(\ref{c5}) tells us there are two consecutive edges in $E_3$, say $e_0$ and $e_1$. Hence $n_5\geq 1$ and $n_4+n_5=2$. We may assume $e_2\in E_4\cup E_5$ and $e_3\in E_5$.
% Suppose $m(e_3)=3$. Then since $m(e_2)\geq 2$, Conf(\ref{c3}) implies $m(e_1)=1$ and $v_0v_1x$ is a tough triangle, contradicting Conf(\ref{c13}).
% Thus $m(e_3)=2$ and $e_3$ is incident with a tough triangle $v_0v_3x$, and so by Conf(\ref{c13}), $m(e_1)=1$. But then by (4) the second region for $e_1$ is a tough triangle, contradicting Conf(\ref{c13}).
Since $m(e_2) \geq 2$, Conf(\ref{c3}) implies that $m(e_1) + m(e_3) \leq 4$. Thus by (4) and (6), either the second region for $e_1$ is a tough triangle, or the second region for $e_3$ is a tough triangle and $m(e_1)=2$. Further, $m^+(e_1) + m^(e_3) = 5$. This contradicts Conf(\ref{c13}).
We conclude that Case 3 does not apply.

\bigskip

This completes the proof of \ref{smallovercharge}.~\bbox

\noindent{\bf Proof of \ref{unav}.\ \ }
Suppose that $(G,m)$ is a prime $7$-target, and let $\alpha, \beta$ be as before. Since the sum over all regions $r$ of
$\alpha(r)+\beta(r)$ is positive, there is a region $r$ with $\alpha(r)+\beta(r)>0$. But this is contrary to one of
\ref{bigovercharge}, \ref{triovercharge1}, \ref{triovercharge2}, \ref{smallovercharge}. This proves \ref{unav}.~\bbox

\section{Reducibility}

Now we begin the second half of the paper, devoted to proving the following.

\begin{thm}\label{reduc}
Every minimum $7$-counterexample is prime.
\end{thm}
Again, the proof is broken into several steps. Clearly no minimum $7$-counterexample $(G,m)$ has an edge $e$ with $m(e) = 0$, because deleting $e$
would give a smaller $7$-counterexample; and by \ref{counterex}, every minimum $7$-counterexample satisfies the conclusions of \ref{counterex}. Thus, it remains to check that 
$(G,m)$ contains none of Conf(1)--Conf(14). In \cite{ces12} we found it was sometimes just as easy to prove a result for general $d$ instead of $d = 8$, and so the following theorem is proved there.

\begin{thm}\label{touchtri}
If $(G,m)$ is a minimum $d$-counterexample, then every triangle has multiplicity less than $d$.
\end{thm}

It turns out that Conf(\ref{c1}) is a reducible configuration for every $d$ as well; this follows easily from \ref{counterex} and is proved in \cite{ces12}.

\begin{thm}\label{cubic}
No minimum $d$-counterexample contains Conf(\ref{c1}).
\end{thm}

% If $C$ is a cycle of length four in $G$, say with vertices $u,v,w,x$ in order, let $m'$ be defined as follows:
% $m'(uv) = m(uv)-1$, $m'(vw) = m(vw)+1$, $m'(wx)= m(wx)-1$, $m'(ux) = m(ux)+1$, and $m'(e) = m(e)$ for all other edges $e$. If $(G,m)$ is a minimum $d$-counterexample, then
% because
% of the second statement of \ref{counterex}, it follows that $(G,m')$ is a $d$-target. (Note that possibly $m'(uv),m'(wx)$ are zero; this is the reason to
% permit $m(e)=0$ in a $d$-target.) We say that $(G,m')$ is obtained from $(G,m)$ by {\em switching on the sequence $u\d v\d w\d x\d u$}.
% If $(G,m')$ is smaller than $(G,m)$, we say that the sequence $u\d v\d w\d x\d u$ is {\em switchable}.

If $(G,m)$ is a $d$-target, and $x,y$ are distinct vertices both incident with some common region $r$, we define $(G,m)+xy$ to be
the $d$-target $(G',m')$ obtained as follows:
\begin{itemize}
\item If $x,y$ are adjacent in $G$, let $(G',m') = (G,m)$.
\item If $x,y$ are non-adjacent in $G$, let $G'$ be obtained from $G$ by adding a new edge $xy$, extending the drawing of $G$ to one of $G'$ and setting $m'(e) = m(e)$
for every $e\in E(G)$ and $m'(xy) = 0$.
\end{itemize}

Let $(G,m)$ be a $d$-target, and let $x\d u\d v\d y$ be a three-edge path of $G$, where $x,y$ are incident with a common region. Let $(G',m')$ be 
obtained as follows:
\begin{itemize}
\item If $x,y$ are adjacent in $G$, let $G' = G$, and otherwise let $G'$ be obtained from $G$ by adding the edge $xy$ and extending the drawing of $G$ to one of $G'$.
\item Let $m'(xu) = m(xu)-1$, $m'(uv) = m(uv)+1$, $m'(vy)= m(vy)-1$, $m'(xy) = m(xy)+1$ if $xy\in E(G)$ and $m'(xy) = 1$ otherwise, and $m'(e) = m(e)$ for all other edges $e$.
\end{itemize}
If $(G,m)$ is a minimum $d$-counterexample, then
because
of the second statement of \ref{counterex}, it follows that $(G',m')$ is a $d$-target. 
We say that $(G',m')$ is obtained from $(G,m)$ by {\em switching on the sequence $x\d u\d v\d y$}.
If $(G',m')$ admits a $d$-edge-colouring, we say that the path $x\d u\d v\d y$ is {\em switchable}.

\begin{thm}\label{dense}
No minimum $7$-counterexample contains Conf(\ref{c2}) or Conf(\ref{c3}).
\end{thm}
\Proof
To handle both cases at once, let us assume that $(G,m)$ is a $7$-target, and $uvw,uwx$ are triangles with $m(uv) + m(uw) + m(vw)+ m(ux)\ge 7$,
(where possibly $m(uw) = 0$); and either $(G,m)$ is a minimum $7$-counterexample, or $m(uw) = 0$ and deleting $uw$ gives a minimum $7$-counterexample $(G_0,m_0)$ say.
Let $(G,m')$ be obtained by switching $(G,m)$ on $u\d v\d w\d x$.
\\
\\
(1) {\em $(G,m')$ is not smaller than $(G,m)$.}
\\
\\
Because suppose it is. Then it admits a $7$-edge-colouring; because if $(G,m)$ is a minimum $7$-counterexample this is clear, and otherwise $m(uw) = 0$, and
$(G',m')$ is smaller than $(G_0,m_0)$. Let $F_1'\l F_7'$ be a $7$-edge-colouring of $(G',m')$.
Since 
$$m'(uv)+m'(uw)+m'(vw)+m'(ux)\ge 8,$$
one of $F_1'\l F_7'$, say $F_1'$, contains two of $uv,uw,vw,ux$ and hence contains $vw,ux$. Then 
$$(F_1'\setminus\{vw,ux\})\cup \{uv,wx\}$$
is a perfect matching, and it together with $F_2'\l F_7'$ provide a $7$-edge-colouring of $(G,m)$, a contradiction. This proves (1).

\bigskip

From (1) we deduce that $\max(m(ux),m(vw))<\max(m(uv),m(wx))$. Consequently, 
$$m(uv) + m(uw) + m(vw)+ m(wx)\le 6,$$ by (1) applied with $u,w$ exchanged;
and 
$$m(uv)+ m(ux)+m(wx) + m(uw)\le 6,$$ by (1) applied with $v,x$ exchanged. Consequently $m(ux)>m(wx)$, and hence $m(ux)\ge 2$; and
$m(vw)>m(wx)$, and so $m(vw)\ge 2$. Since $m(uv) + m(uw) + m(vw)+ m(wx)\le 6$ and $m(vw)\ge 2$, it follows that $m(uv)\le 3$; and since
$\max(m(ux),m(vw))<\max(m(uv),m(wx))$, it follows that $m(uv) = 3$, $m(vw)=m(ux) = 2$ and $m(wx) = 1$. But this is contrary to (1), and so
proves \ref{dense}.~\bbox

\section{Guenin's cuts}

Next we introduce a method of Guenin~\cite{guenin}.
Let $G$ be a three-connected graph drawn in the plane, and let $G^*$ be its dual graph; let us identify $E(G^*)$ with $E(G)$ in the natural way.
A {\em cocycle} means the edge-set of a cycle of the dual graph; thus, $Q\subseteq E(G)$ is a cocycle of $G$
if and only if $Q$ can be numbered $\{e_1\l e_k\}$ for some $k\ge 3$ and there are distinct regions $r_1\l r_k$ of $G$ such that
$1\le i\le k$, $e_i$ is incident
with $r_i$ and with $r_{i+1}$ (where $r_{k+1}$ means $r_1$). 
Guenin's method is the use of the following theorem, a proof of which is given in \cite{ces12}.

\begin{thm}\label{cuts}
Suppose that $d\ge 1$ is an integer such that every $(d-1)$-regular oddly $(d-1)$-edge-connected planar graph is $(d-1)$-edge-colourable.
Let $(G,m)$ be a minimum $d$-counterexample, and let $x\d u\d v\d y$ be a path of $G$ with $x,y$ on a common region. 
Let $(G',m')$ be obtained by switching on $x\d u\d v\d y$, and
let $F_1\l F_d$ be a $d$-edge-colouring of $(G',m')$, where $xy\in F_k$. 
Then none of $F_1\l F_d$ contain both $uv$ and $xy$.
Moreover, let $I = \{1\l d\}\setminus \{k\}$ if $xy\notin E(G)$, and $I = \{1\l d\}$ if $xy\in E(G)$.
Then for each $i\in I$,  there is a cocycle $Q_i$ of $G'$ with the following properties:
\begin{itemize}
\item for $1\le j\le d$ with $j \ne i$, $|F_j\cap Q_i| = 1$;
\item $|F_i\cap Q_i|\ge 5$;
\item there is a set $X\subseteq V(G)$ with $|X|$ odd such that $\delta_{G'}(X) = Q_i$; and
\item $uv,xy\in Q_i$ and $ux,vy\notin Q_i$.
\end{itemize}
\end{thm}

\bigskip
By the result of~\cite{dvorak}, every $6$-regular oddly $6$-edge-connected planar graph is $6$-edge-colourable, so we can apply \ref{cuts} when $d = 7$.

\begin{thm}\label{squareopp}
No minimum $7$-counterexample contains Conf(\ref{c4}) or Conf(\ref{c5}).
\end{thm}
\Proof
To handle both at once, let us assume that $(G,m)$ is a $7$-target, and $uvw,uwx$ are two triangles with $m^+(uv)+m(uw)+m^+(wx)\ge 6$; and either $(G,m)$ is a 
minimum $7$-counterexample, or
$m(uw) = 0$ and deleting $uw$ gives a minimum $7$-counterexample. 
We claim that $u\d x\d w\d v\d u$ is switchable. For suppose not; then we may assume that $m(vw)>max(m(uv),m(wx))$ and $m(vw)\ge m(ux)$. Now we do not have Conf(\ref{c2}) or Conf(\ref{c3}) by \ref{dense} so
$$m(uv)+m(uw)+m(vw)+m(wx)\le 6,$$
and yet $m(uv)+m(uw)+m(wx) \ge 4$ since $m^+(uv)+m(uw)+m^+(wx)\ge 6$; and so $m(vw)\le 2$. Consequently $m(uv),m(wx) = 1$, and $m(ux)\le 2$. Since
$u\d x\d w\d v\d u$ is not switchable, it follows that $m(ux)=m(vw)= 2$; and since $m^+(uv)+m(uw)+m^+(wx)\ge 6$, it follows that $m(uw)\ge 2$ giving Conf(\ref{c2}), contrary to \ref{dense}.
This proves that $u\d x\d w\d v\d u$ is switchable.

Let $r_1,r_2$ be the second regions incident with $uv,wx$ respectively, and for $i = 1,2$ let $D_i$ be the set of doors for $r_i$.
Let $k = m(uv)+m(uw)+m(wx)+2$. Let $(G,m')$ be obtained by switching on $u\d x\d w\d v\d u$, and let $F_1\l F_7$ be a $7$-edge-colouring of $(G,m')$, where $F_i$ contains one of
$uv,uw,wx$ for $1\le i\le k$. For $1\le i\le 7$, let $Q_i$ be as in \ref{cuts}. 
\\
\\
(1) {\em For $1\le i\le 7$, either $F_i\cap Q_i\cap D_1\ne \emptyset$, or $F_i\cap Q_i\cap D_2\ne \emptyset$; and both are nonempty if either $k = 7$ or $i=7$.}
\\
\\
For let the edges of $Q_i$ in order be $e_1\l e_n,e_1$,
where $e_1 = wx$, $e_2 = uw$, and $e_3 =uv$. Since
$F_j$ contains one of $e_1,e_2,e_3$ for $1\le j\le k$, it follows that none of $e_4\l e_n$ belongs to any $F_j$ with $j\le k$ and $j\ne i$, and, 
if $k=6$ and $i\ne 7$, that only one of them is in $F_7$. But since at most one of $e_1,e_2,e_3$ is in $F_i$ and $|F_i\cap Q_i|\ge 5$, it follows that $n\ge 7$; so
either $e_4,e_5$ belong only to $F_i$, or $e_n, e_{n-1}$ belong only to $F_i$, and both if $k = 7$ or $i = 7$. But if $e_4,e_5$ are only contained in $F_i$, then
they both have multiplicity one, and are disjoint, so $e_4$ is a door for $r_1$ and hence $e_4\in F_i\cap Q_i\cap D_1$. Similarly if $e_n,e_{n-1}$ are only
contained in $F_i$ then $e_n\in F_i\cap Q_i\cap D_2$. This proves (1).

\bigskip

Now $k\le 7$, so one of $r_1,r_2$ is small since $m^+(uv)+m(uw)+m^+(wx)\ge 6$; and if $k = 7$ then by (1) $|D_1|,|D_2|\ge 7$, a contradiction. Thus $k =6$, 
so both $r_1,r_2$ are small, but from (1) $|D_1|+|D_2|\ge 8$, again a contradiction. 
This proves \ref{squareopp}.~\bbox

\begin{thm}\label{bigtri}
No minimum $7$-counterexample contains Conf(\ref{c6}).
\end{thm}
\Proof
Let $(G,m)$ be a minimum $7$-counterexample, and suppose that $uvw$ is a triangle
with $m^+(uv)+m^+(uw) = 6$ and either $m(uv)\geq 3$ or $m(uv)=m(vw)=m(uw)=2$ or $u$ has degree at least four. Let $r_1,r_2$ be the second regions for $uv,uw$ respectively, and for $i = 1,2$
let $D_i$ be the set of doors for $r_i$. Since we do not have Conf(\ref{c4}) by \ref{squareopp}, neither of $r_1,r_2$ is a triangle. 
Let $tu$ be the edge incident with $r_2$ and $u$ different from $uv$. It follows from \ref{cubic} that we do not have Conf(\ref{c1}) so
$m(tu)\le 2$, since $m(uv)+m(uw)\ge 4$ and $m(vw)+max(m(uv),m(uw))\geq 4$. By \ref{touchtri}, $m(vw)\le m(uv)$. Thus the path $t\d u\d v\d w$ is switchable. Note that $t,w$ are non-adjacent in $G$, since $r_2$
is not a triangle.

Let $(G',m')$ be obtained by switching on this path, and
let $F_1\l F_7$ be a $7$-edge-colouring of it. Let $k = m(uv)+m(uw)+2$; thus $k\ge 6$, since $m(uv)+m(uw)\ge 4$. By \ref{cuts} we may assume that for $1\le j<k$, $F_j$ contains one of
$uv,uw$, and $tw\in F_k$. 

Let $I = \{1\l 7\}\setminus\{k\}$, and for each $i\in I$, let $Q_i$ be as in \ref{cuts}. Now let $i\in I$, and let the edges of $Q_i$ in order be $e_1\l e_n,e_1$, where $e_1 = uv$, 
$e_2 = uw$, and $e_3 = tw$. Since $F_j$ contains one of $e_1,e_2,e_3$ for $1\le j\le k$ it follows that none of $e_4\l e_n$ belong to any $F_j$ with $j\le k$; and if
$k = 6$ and $i\ne 7$, only one of them belongs to $F_7$. Since $F_i$ contains at most one of $e_1,e_2,e_3$ and $|F_i\cap Q_i|\ge 5$, it follows that $n\ge 7$, and so
either $e_4,e_5$ belong only to $F_i$, or $e_n, e_{n-1}$ belong only to $F_i$; and both if either $k = 7$ or $i = 7$. 
Thus either $e_4\in F_i\cap Q_i\cap D_2$ or $e_n\in F_i\cap Q_i\cap D_1$, and both if $k = 7$ or $i = 7$. 
Since $k\le 7$, 
one of $r_1,r_2$ is small since $m^+(uv)+m^+(uw) = 6$; and yet if $k = 7$ then $|D_1|,|D_2| \ge  |I| =  6$, a contradiction. Thus $k=6$, so $r_1,r_2$ are both small, and yet
$|D_1|+|D_2|\ge 7$, a contradiction.
This proves \ref{bigtri}.~\bbox

\begin{thm}\label{3heavy}
No minimum $7$-counterexample contains Conf(\ref{c7}).
\end{thm}

\Proof
Let $(G,m)$ be a minimum $7$-counterexample, with an edge $uv$ with $m^+(uv)\geq 4$ incident with regions $r_1$ and $r_2$ and $r_1$ has length at least four. Suppose further that every edge $e$ of $C_{r_1}$ disjoint from $uv$ is $2$-heavy and not incident with a triangle with multiplicity three. It is enough to show that there are at least four edges on $C_{r_1}$ disjoint from $uv$ that are not $3$-heavy. By \ref{sixdoors} and \ref{fardoor} we do not have Conf(\ref{c11}) or Conf(\ref{c9}). Hence $m(uv)=3$ and $r_2$ is small.

Let $x\d u\d v\d y$ be a path of $C_r$. By \ref{squareopp} we do not have Conf(\ref{c5}), so $x$ and $y$ are not adjacent in $G$. Since $G$ has minimum degree three, $m(uv)\geq m(ux),m(vy)$ so $x\d u\d v\d y$ is switchable; let $(G',m')$ be obtained from $(G,m)$ by switching on it, and let $F_1\l F_7$ be a $7$-edge-colouring of $(G',m')$. 
% Let $k=m'(uv) + m'(xy)=5$. Since $Q_i$ contains both $uv$ and $xy$ for each $i\in I$, it follows that for $1\le j\le 7$, $F_j$ contains at most one
% of $uv,xy$. 

% For $i\in I$, let $Q_i$ be as in \ref{cuts}. 
Since $m'(uv) + m'(xy) =5$ we may assume by \ref{cuts} that $uv\in F_i$ for $1\le i \le 4$ and $xy\in F_5$.
Let $I = \{1\l 7\}\setminus \{5\}$ and for $i\in I$, let the edges of $Q_i$ in order be $e_1^i\l e_n^i,e_1^i$, where $e_1^i = uv$ and $e_2^i = xy$. 

Since $|F_i\cap Q_i|\ge 5$ and $F_i$ contains at most one of
$e_1^i,e_2^i$, it follows that $n\ge 6$.
Let $D_2$ denote the set of doors for $r_2$.
\\
\\
(1) {\em Let $i\in I$. If $i>k$ then $F_i\cap D_2$ is nonempty. Further, if $F_i\cap D_2$ is empty, or $i>k$ then $e_3^i$ is not $3$-heavy, and either
\begin{itemize} 
  \item $e_3^i$ belongs to $F_i$, or
  \item $e_4^i$ belongs to $F_i$ and $m(e_3^i)=m(e_4^i)=1$ and $e_3^i, e_4^i$ belong to a triangle.
\end{itemize} } 
For $1\le j\le 5$, $F_j$ contains one of $e_1^i,e_2^i$; and hence $e_3^i\l e_n^i\notin F_j$ for all $j\in \{1\l 5\}$ with $j\ne i$. Therefore $e_3^i\l e_n^i $ belong only to $ F_i, F_6, F_7$. Since $e_3^6$ is $2$-heavy, one of $e_3^6,e_4^6$ does not belong to $F_6$ and therefore belongs to $F_7$. It follows that $e_n^6, e_{n-1}^6\notin F_7$ so $F_6\cap D_2$ is nonempty. Similarly, $F_7\cap D_2$ is nonempty. This proves the first assertion.

Suppose $F_i\cap D_2$ is empty, or $i>5$; we have $|\{e_n^i,e_{n-1}^i\}\cap (F_6\cup F_7)|\geq 1$. Without loss of generality say $|\{e_n^i,e_{n-1}^i\}\cap F_6|\geq 1$. It follows that $e_3^i$, $e_4^i$ belong only to $F_i, F_7$, so $e_3^i$ is not $3$-heavy. 
% Since each $F_i$ is a perfect matching and $e_3^i\l e_n^i \in F_i\cup F_7$, $e_3^i$ cannot be $3$-heavy. 
On the other hand, $e_3^i$ is $2$-heavy by hypothesis, so if $e_3^i\notin F_i$, then $e_3^i$ has multiplicity one, $e_3^i\in F_7$, $e_4^i$ belongs to $F_i$, has multiplicity one. Since $e_3^i$ is $2$-heavy, $e_3^i$ and $e_4^i$ belong to a triangle.
This proves (1).

Let $I_1$ denote the indices $i\leq 6, i\neq 5$ such that $e_3^i$ is not $3$-heavy and either $e_3^i\in F_i$, or $e_4^i\in F_i$ and $e_3^i,e_4^i$ have multiplicity one and belong to a triangle incident with $r_1$.
From (1) and because $r_2$ is small, $|I_1|\geq 4$. 
Suppose that for $i\neq i' \in I_1$, the corresponding edges $e_3^i$ and $e_3^{i'}$ are the same. We may assume $i'\leq 4$. If $e_3^i\in F_{i'}$, this is a contradiction. Otherwise $m(e_3^i)= m(e_4^i) = 1$ and $e_3^i, e_4^i$ belong to a triangle incident with $r_1$. It follows that $e_4^i = e_4^{i'}$ since $e_3^i$ is not incident with a triangle of multiplicity three, and so $e_4^i\in F_{i'}$, a contradiction.

% Further, since no edge $e_3^i$ is incident with a triangle of multiplicity three, for a pair of such indices $i,i'<k$, the corresponding edges $e_3^i$ and $e_3^{i'}$ are distinct, and are distinct from $e_3^6$.
It follows that there are at least four edges of $C_r$ disjoint from $uv$ that are not $3$-heavy.
This proves \ref{3heavy}.~\bbox

\begin{thm}\label{sevengon}
No minimum $7$-counterexample contains Conf(\ref{c8}).
\end{thm}
\Proof
%A region $r$ of length at least four, an edge $e$ of $C_r$ with $m^+(e)=m(e)+1=4$, an edge $f$ disjoint from $e$ with $m^+(f)=m(f)+1=2$, where every edge of $C_r\setminus\{f\}$ disjoint from $e$ is $3$-heavy.
Let $(G,m)$ be a minimum $7$-counterexample, with an edge $uv$ with multiplicity three, incident with regions $r$ and $r_1$ where $r_1$ is small. Suppose there is an edge $f$ disjoint from $e$ with $m^+(f)=m(f)+1=2$, where every edge of $C_r\setminus\{f\}$ disjoint from $e$ is $3$-heavy with multiplicity at least two. Since $e$ and $f$ are disjoint $r$ has length at least four. Let $x\d u\d v\d y$ be a path of $C_r$. By \ref{squareopp} we do not have Conf(\ref{c5}), so $x$ and $y$ are not adjacent in $G$.
Since $G$ has minimum degree at least three, it follows that $m(uv)\geq m(ux),m(vy)$ so $x\d u\d v\d y$ is switchable; let $(G',m')$ be obtained from $(G,m)$ by switching on it, and let $F_1\l F_7$ be a $7$-edge-colouring of $(G',m')$.  Since $m'(uv) + m'(xy)=5$ we may assume by \ref{cuts} that $uv\in F_i$ for $1\le i \le 4$ and $xy\in F_5$.
Let $I = \{1\l 7\}\setminus \{5\}$ and for $i\in I$, let $Q_i$ be as in \ref{cuts}. 

For $i\in I$, let the edges of $Q_i$ in order be $e_1\l e_n,e_1$, where $e_1 = uv$ and $e_2 = xy$. Since $|F_i\cap Q_i|\ge 5$ and $F_i$ contains at most one of
$e_1,e_2$, it follows that $n\ge 6$.
For $1\le j \le 5$, $F_j$ contains one of $e_1,e_2$; and hence for all $j\in \{1\l 5\}$, $e_3\l e_n\notin F_i$, and so $e_3\l e_n$ belong only to $F_i, F_6$ or $F_7$. 
In particular when $i\in \{6,7\}$, $e_3$ is not $3$-heavy and so $e_3=f$. It follows $f$ belongs only to $F_6,F_7$; assume without loss of generality $f\in F_6$.
Let $D_1$ denote the set of doors for $r_1$. Denote by $r_2$ the second region for $f$ and by $D_2$ its set of doors.
\\
\\
(1) {\em Let $i\in I$. At least one of $F_i\cap Q_i \cap D_1$, $F_i\cap Q_i \cap D_2$ is nonempty, and if $i=7$ then both are nonempty.}
\\
\\
Suppose $i=7$. Then $e_3=f\in F_6$ and $e_4\l e_n $ belong only to $F_7$, and so $e_4$ is a door for $r_2$ and $e_n$ is a door for $r_1$.
Now suppose $i<7$. 
If $e_3=f$, then since $F_i$ contains at most one of
$e_1,e_2,e_3$ and $|F_i\cap Q_i|\ge 5$, it follows that $n\ge 7$.
It follows that $e_4\l e_n $ belong only to $F_7$ or $F_i$, and so either $e_4$ is a door for $r_2$ or $e_n$ is a door for $r_1$ as required.
If $e_3\neq f$ then $e_3$ is $3$-heavy, and so $F_i,F_6,F_7$ each contain one of $e_3,e_4$. Therefore $e_{n-1}, e_n$ belong only to $F_i$, and so $e_n$ is a door for $r_1$.
This proves (1). \\

By (1), $|D_1| + |D_2| \geq 7$, but $r_1$ and $r_2$ are both small, a contradiction.
This proves \ref{sevengon}.~\bbox

\begin{thm}\label{fardoor}
No minimum $7$-counterexample contains Conf(\ref{c9}).
\end{thm}
\Proof
Let $(G,m)$ be a minimum $7$-counterexample, and suppose that some edge $uv$ with $m(uv)=4$
is incident with a region $r$ of length at least four. 
Let $x\d u\d v\d y$ be a path of $C_{r_1}$. 
If $x$ and $y$ are adjacent, then since we do not have Conf(\ref{c5}) by \ref{squareopp}, $xy$ is incident with a big region. Therefore may assume $x$ and $y$ are nonadjacent.

We will show $r$ has a door $f$ disjoint from $uv$, and that if $m(xu)\geq 2$ then $f$ is also disjoint from $xu$ (and similarly for $vy$.)

Since $m(e)\ge 4$, this path is
switchable; let $(G',m')$ be obtained from $(G,m)$ by switching on it, and let $F_1\l F_7$ be a $7$-edge-colouring of $(G',m')$. %Let $k = m'(uv)+m'(xy)= 6$.

Thus
we may assume that $uv\in F_i$ for $1\le i\le 5$, and $xy\in F_6$. Further, if $m(xu)\geq 2$ then $xu\in F_7$ and simlarly for $vy$.
Let $I = \{1\l 7\}\setminus \{6\}$.
For $i\in I$, let $Q_i$ be as in \ref{cuts}. Since $Q_i$ contains both $uv,xy$ for each $i\in I$, it follows that for $1\le j\le 7$, $F_j$ contains at most one
of $uv,xy$. 

Consider now $Q_7$, and let the edges of $Q_7$ in order be $e_1 \l e_n, e_1$ where $e_1=uv$ and $e_2=xy$. For $1\leq j \leq 6$, $F_j$ contains one of $e_1,e_2$, and hence $e_3\l e_n $ belong only to $F_7$. Since $e_3 \in C_r\setminus \{xu,uv,vy\}$ by the choice of the switchable path, $e_3$ is a door for $r$ disjoint from $uv$. Further if $m(xu) \geq 2$ then $e_3$ is disjoint from $xu$, and similarly for $vy$.

This proves \ref{fardoor}.~\bbox

\begin{thm}\label{4gon2heavy}
No minimum $7$-counterexample contains Conf(\ref{c10}).
\end{thm}
\Proof
Let $(G,m)$ be a minimum $7$-counterexample, and suppose that there is a region $r$ of length between four and six incident with an edge $uv$ with multiplicity four, and suppose that $m^+(e)\geq 2$ for every edge $e$ of $C_r$ disjoint from $uv$. Let $x\d u\d v\d y$ be a path of $C_r$. By \ref{squareopp}, we do not have Conf(\ref{c5}) so $x$ and $y$ are not adjacent in $G$ (and $r$ has length five or six). 
Since $m(uv)=4$, the path $x\d u\d v\d y$ is switchable; let $(G',m')$ be obtained from $(G,m)$ by switching on it, and let $F_1\l F_7$ be a $7$-edge-colouring of $(G',m')$. 
By \ref{cuts}
we may assume that $uv\in F_i$ for $1\le i\le 5$, and $xy\in F_6$.
Let $I = \{1\l 7\}\setminus \{6\}$ and for $i\in I$, let $Q_i$ be as in \ref{cuts}. 

Define $\ell =|F_7 \cap E(C_r)\setminus\{xu,uv,vy\}|$. Suppose $\ell=0$; then let the edges of $Q_7$ in order be $e_1\l e_n,e_1$, where $e_1 = uv$ and $e_2 = xy$. Since $|F_i\cap Q_i|\ge 5$ and $F_i$ contains at most one of
$e_1,e_2$, it follows that $n\ge 6$. For $1\le j \le 6$, $F_j$ contains one of $e_1,e_2$; and hence $e_3\l e_n$ belong only to $F_7$. But $e_3$ is an edge of $E(C_r)\setminus\{xu,uv,vy\}$ by the choice of the switchable path, a contradiction. Thus $\ell \geq 1$.
Fix an edge $f\in F_7 \cap E(C_r)\setminus\{xu,uv,vy\}$ and let $I_1$ denote the indices $i\in I$ for which $f\in Q_i$.
\\
\\
(1) $|I_1|\leq 3$.
\\
\\
% Let $f\in F_7 \cap E(C_r)\setminus\{xu,uv,vy\}$, d
Denote by $r_2$ the second region for $f$ and denote by $D_2$ the set of doors for $r_2$. Suppose that $|I_1|\geq 4$.
For $i\in I_1$, let the edges of $Q_i$ in order be $e_1\l e_n,e_1$, where $e_1 = uv$, $e_2 = xy$ and $e_3=f$. Since $|F_i\cap Q_i|\ge 5$ and $F_i$ contains at most one of
$e_1,e_2$,$e_3$, it follows that $n\ge 7$. For $1\le j \le 7$, $F_j$ contains one of $e_1,e_2,e_3$; and hence $e_4\l e_n$ belong only to $F_i$.
Further, $e_4$ is incident with $r_2$ and therefore is a door for $r_2$. But then $|D_2|\geq 4$, so $m^+(f)=1$, a contradiction. This proves (1).

Since $r$ has length at most six, there are two cases:

\noindent{\bf Case 1:} $\ell=1$.
Let $f\in F_7 \cap E(C_r)\setminus\{xu,uv,vy\}$, denote by $r_2$ the second region for $f$ and denote by $D_2$ the set of doors for $r_2$. Since the edges of $C_r\setminus \{xu,uv,vy,f\}$ each belong to $F_j$ for some $j\neq 7$, there are at most two indices $i\in I$ for which $f\notin Q_i$. 
But then we have $|I_1|\geq 4$, contradicting (1).

\noindent{\bf Case 2:} $\ell=2$.
Let $f,f' \in F_7 \cap E(C_r)\setminus\{xu,uv,vy\} $. If $m(f') \geq 2$, then $f'\in F_j$ for some $j\neq 7$, and so there are at most two values of $i\in I$ for which $f\notin Q_i$. Then $|I_1|\geq 4$, contradicting (1). So $m(f')=1$ and by symmetry, $m(f)=1$. 
There is at most one value of $i\in I$ for which $f,f'\notin Q_i$. Therefore, without loss of generality we may assume there at least three indices $i\in I$, $f\in Q_i$, and so $|I_1|=3$. Denote by $r_2$ the second region for $f$ and $D_2$ the set of doors for $r_2$. For each $i\in I_1$, it follows that $e_4\l e_n $ belong only to $F_i$, and $e_4$ is incident with $r_2$ and therefore is a door for $r_2$. 
Further, since $f$ and $f'$ are disjoint and have multiplicity one, $f$ is a door for $r_2$. If follows that $|D_2|\geq 4$, so $m^+(f)=1$, a contradiction.

This completes the proof of \ref{4gon2heavy}.~\bbox

\begin{thm}\label{sixdoors}
No minimum $7$-counterexample contains Conf(\ref{c11}).
\end{thm}
\Proof
Let $(G,m)$ be a minimum $7$-counterexample, and suppose that some edge $uv$
is incident with regions $r_1,r_2$ where either $m(uv)=4$ and $r_2$ is small, or $m(uv)\ge 5$.
By exchanging $r_1,r_2$ if necessary, we may assume that if $r_1,r_2$ are both small, then the 
length of $r_1$ is at least the length of $r_2$. 
Suppose $r_1$ is a triangle. Then by \ref{dense} we do not have Conf(\ref{c3}), and so $r_2$ is not a triangle and therefore $r_2$ is big. Then by hypothesis, $m(uv)\ge 5$, contradicting \ref{touchtri}. Thus $r_1$ is not a triangle.

Let $x\d u\d v\d y$ be a path of $C_{r_1}$. By \ref{squareopp} we do not have Conf(\ref{c5}) so $x,y$ are non-adjacent in $G$.
Since $m(e)\ge 4$, this path is
switchable; let $(G',m')$ be obtained from $(G,m)$ by switching on it, and let $F_1\l F_7$ be a $7$-edge-colouring of $(G',m')$. 
Let $k = m(uv)+2\ge 6$. By \ref{cuts} we may assume that $uv\in F_i$ for $1\le i\le k-1$, and $xy\in F_k$, and so $k\le 7$.
Let $I = \{1\l 7\}\setminus \{k\}$ and for $i\in I$, let $Q_i$ be as in \ref{cuts}. 

Let $D_1$ be the set of doors for $r_1$ that are disjoint from $e$, and let $D_2$ be the set of doors for $r_2$.
\\
\\
(1) {\em For each $i\in I$, one of $F_i\cap Q_i\cap D_1, F_i\cap Q_i\cap D_2$ is nonempty, and if $k=7$ or $i>k$ then both are nonempty.}
\\
\\
Let $i\in I$, and let the edges of $Q_i$ in order be $e_1\l e_n,e_1$, where $e_1 = uv$ and $e_2 = xy$. Since $|F_i\cap Q_i|\ge 5$ and $F_i$ contains at most one of
$e_1,e_2$, it follows that $n\ge 6$.
Suppose that $k = 7$. Then for $1\le j\le 7$, $F_j$ contains one of $e_1,e_2$; and hence $e_3\l e_n\notin F_j$ for all $j\in \{1\l 7\}$ with $j\ne i$. It follows
that $e_n,e_{n-1}$ belong only to $F_i$ and hence $e_n\in F_i\cap Q_i\cap D_2$. Since this holds for all $i\in I$, it follows that $|D_2|\ge |I|\ge 6$. Hence $r_2$
is big, and so by hypothesis, $m(uv)\ge 5$. Since $xy\notin E(G)$, $e_3$ is an edge of $C_{r_1}$, and since $e_3,e_4$ belong
only to $F_i$, it follows that $e_3$ is a door for $r_1$. But $e_3\ne ux,vy$ from the choice of the switchable path, and so $e_3\in F_i\cap Q_i\cap D_1$. 
Hence in this case (1) holds.

Thus we may assume that $k = 6$ and so $I = \{1\l 5,7\}$; we have $m(e) = 4$, and $r_2$ is small, and 
$uv\in F_1\l F_5$, and $xy\in F_6$.
If $i=7$, then since $uv,xy\in Q_i$ and $F_j$ contains one of $e_1,e_2$ for all $j\in \{1\l 6\}$, it follows as before that $e_3\in F_i\cap Q_i\cap D_1$ and 
$e_n\in F_i\cap Q_i\cap D_2$.
We may therefore assume that $i\le 6$. For $1\le j\le 7$
with $j\ne i$, $|F_j\cap Q_i| = 1$, and for $1\le j\le 6$, $F_j$ contains one of $e_1,e_2$. Hence $e_3\l e_n$ belong only to $F_i$ and to $F_7$, and only
one of them belongs to $F_7$. If neither of $e_n,e_{n-1}$ belong to $F_7$ then $e_n\in F_i\cap Q_i\cap D_2$ as required; so we assume that $F_7$ contains one
of $e_n,e_{n-1}$; and so $e_3\l e_{n-2}$ belong only to $F_i$. Since $n\ge 6$, it follows that $e_3\in F_i\cap Q_i\cap D_1$ as required. This proves (1).

\bigskip

If $k = 7$, then (1) implies that $|D_1|,|D_2|\ge 6$ as required. So we may assume that $k = 6$ and hence $m(e) = 4$ and $xy\notin E(G)$; and $r_2$ is small.
Suppose that there are three values of $i\in \{1\l 5\}$ such that $|F_i\cap D_1| = 1$ and $F_i\cap D_2 = \emptyset$,
say $i = 1,2,3$. Let $f_i\in F_i\cap D_1$ for $i = 1,2,3$, and we may assume that $f_3$ is between $f_1$ and $f_2$
in the path $C_{r_1}\setminus\{uv\}$. Choose $X\subseteq V(G')$ such that $\delta_{G'}(X) = Q_{3}$. Since only one edge of $C_{r_1}\setminus\{e\}$
belongs to $Q_3$, one of $f_1, f_2$ has both ends in $X$ and the other has both ends in $V(G')\setminus X$; say $f_1$ has both ends in $X$.
Let $Z$ be the set of edges with both ends in $X$. Thus $(F_1\cap Z)\cup (F_2\setminus Z)$ is a perfect matching, since  $e\in F_1\cap F_2$, and no other edge
of $\delta_{G'}(X)$ belongs to $F_1\cup F_2$; and similarly $(F_2\cap Z)\cup (F_1\setminus Z)$ is a perfect matching. Call them $F_1', F_2'$ respectively.
Then $F_1', F_2',F_3, F_4\l F_7$ form a $7$-edge-colouring of $(G', m')$, yet the only edges of $D_1\cup D_2$ included in $F_1'\cup F_2'$ are $f_1,f_2$,
and neither of them is in $F_2'$, contrary to (1).
Thus there are no three such values of $i$; and similarly there are at most two such that $|F_i\cap D_2| = 1$ and $F_i\cap D_1 = \emptyset$.
Thus there are at least two values of $i\in I$ such that $|F_i\cap D_1|+|F_i\cap D_2|\ge 2$ (counting $i = 7$), and so $|D_1|+|D_2|\ge 8$.
But $|D_2|\le 3$ since $r_2$ is small, so $|D_1|\geq 5$.
This proves \ref{sixdoors}.~\bbox

\begin{thm}\label{sixdoors2}
No minimum $7$-counterexample contains Conf(\ref{c12}).
\end{thm}
\Proof
%A region $r$, an edge $uv$ of $C_r$, and a triangle $uvw$ such that $m(uv)=3, m(vw)=2$ and $m(uw)=1$ and at most five edges of $C_r$ (disjoint from $e$) are doors for $r$.
Let $(G,m)$ be a minimum $7$-counterexample, and suppose that some edge $uv$ is incident with a triangle $uvw$ with $m(uv)+m(vw)=5$, and suppose that $uv$ is also incident with a region $r_1$ that has at most five doors disjoint from $v$. Let $tv$ be the edge incident with $r_1$ and $v$ different from $uv$. By \ref{cubic}, we do not have Conf(\ref{c1}) so $m(tv)=1$, and by \ref{touchtri}, $m(uw)=1$. By \ref{dense} we do not have Conf(\ref{c3}), $u$ and $t$ are nonadjacent in $G$. It follows that the path $u \d w \d v \d t$ is switchable; let $(G',m')$ be obtained from $(G,m)$ by switching on it, and let $F_1\l F_7$ be a $7$-edge-colouring of $(G',m')$. 
Since $m'(uv) + m'(uw) + m'(ut) = 7$, we may assume by \ref{cuts} that $ut\in F_7$, and $F_j$ contains one of $uv, vw$ for $1\leq j\le 6$
Let $I = \{1\l 6\}$ and for $i\in I$, let $Q_i$ be as in \ref{cuts}. 
% Since $Q_i$ contains $uv,vw,ut$ for each $i\in I$, it follows that for $1\le j\le 7$, $F_j$ contains at most one
% of $uv,vw,ut$. Thus
% we may assume that $ut\in F_7$, and $F_j$ contains one of $uv, vw$ for $1\leq j\le 6$.

Let $D_1$ be the set of doors for $r_1$ that are disjoint from $v$.
Let $i\in I$, and let the edges of $Q_i$ in order be $e_1\l e_n,e_1$, where $e_1 = vw, e_2 = uv$ and $e_3 = ut$. Since $|F_i\cap Q_i|\ge 5$ and $F_i$ contains at most one of
$e_1,e_2,e_3$, it follows that $n\ge 7$.
For $1\le j\le 7$, $F_j$ contains one of $e_1,e_2,e_3$; and hence $e_3\l e_n\notin F_j$ for all $j\in \{1\l 7\}$ with $j\ne i$. It follows
that $e_4,e_5$ belong only to $F_i$. By the choice of the switchable path $e_4\neq tv$ and hence $e_4\in F_i\cap Q_i\cap D_1$. Since this holds for all $i\in I$, it follows that $|D_1|\ge|I|\ge 6$, a contradiction. 
This proves \ref{sixdoors2}.~\bbox

\begin{thm}\label{switchconf}
Let $(G,m)$ be a minimum $7$-counterexample, let $x\d u\d v\d y$ be a three-edge path of $G$, and let $(G,m')$ obtained by switching on $x\d u\d v\d y$.
If $(G,m)$ is not smaller than $(G,m')$, and $(G,m')$ contains one of Conf(\ref{c1})--Conf(\ref{c12}) then $x\d u\d v\d y$ is switchable.
\end{thm}
\Proof
Suppose that $x\d u\d v\d y$ is not switchable. Then, since $(G,m')$ is a $7$-counterexample and $(G,m)$ is not smaller than $(G,m')$, the latter is a minimum counterexample. But by \ref{cubic}--\ref{sixdoors2}, no minimum $7$-counterexample contains any of Conf(\ref{c1})--Conf(\ref{c12}), a contradiction.
This proves \ref{switchconf}.~\bbox

\begin{thm}\label{newsquare}
No minimum $7$-counterexample contains Conf(\ref{c13}).
\end{thm}
\Proof
Let $(G,m)$ be a minimum $7$-counterexample, with a square $xuvy$ and a tough triangle $uvz$, where $m(uv)+m^+(xy)\geq 4$ and $m(xy)\geq 2$. Since $(G,m)$ does not contain Conf(\ref{c5}) by \ref{squareopp}, we have $m(uv)+m^+(xy) = 4$.
Suppose $m(uv)\geq 3$; then since $xuvy$ is small and $(G,m)$ does not contain Conf(\ref{c6}) by \ref{bigtri}, we have $m(uv)=3$ and $m^+(uz)=m^+(vz)=1$, contradicting the fact that $uvz$ is tough.
Thus $m(uv) \leq 2$.

Since $(G,m)$ does not contain Conf(\ref{c3}) by \ref{dense}, it follows that $m(ux) + m(vy) \leq 4$. Thus the cycle $x\d u\d v \d y \d x$ is switchable; let $(G,m')$ be obtained from $(G,m)$ by switching on it, and let $F_1\l F_7$ be a $7$-edge-colouring of $(G',m')$. 
Let $k = m'(uv) + m'(xy) \in \{5,6\}$.
By \ref{cuts}
we may assume that $uv\in F_i$ for $1\le i\le m'(uv)$, and $xy\in F_i$ for $m'(uv)<i\leq k$.
Let $I = \{1\l 7\}$ and for $i\in I$, let $Q_i$ be as in \ref{cuts}. 
Denote by $r_1$, $r_2$, the second regions for $vz,xy$, respectively, and by $D_1,D_2$ their respective sets of doors.
\\
\\
(1) {\em One of $m^+(uz), m^+(vz) =1$.}
\\
\\
Let $i\in I$, and let the edges of $Q_i$ in order be $e_1^i\l e_{n_i}^i,e_1^i$, where $e_1^i = uv$, $e_2^i = xy$ and $e_{n_i}^i \in \{uz,vz\}$. Since $|F_i\cap Q_i|\ge 5$ and $F_i$ contains at most one of
$e_1^i,e_2^i$, it follows that $n_i\ge 6$.
For $1\le j\le k$, $F_j$ contains one of $e_1^i,e_2^i$; and hence $e_3^i\l e_{n_i}^i\notin F_j$ for all $j\in \{1\l k\}$ with $j\ne i$. 

Suppose $k=6$. We may assume by symmetry that $vz \in Q_7$, and so $m(vz)=1$ and $vz\in F_7$. Also, $uz\in F_i$ for some $m'(uv)<i\leq k$, say $uz\in F_6$.
Let $i\in I\setminus \{6,7\}$. Then since $uz$ and $xy$ both belong to $F_6$, $vz \in Q_i$.
Then since $e_{n_i}^i=vz$ and $vz \notin F_i$, we have $n_i\geq 7$ and $e_3^i\l e_{n_i-1}^i$ belong only to $F_i$. It follows that $F_i\cap Q_i \cap D_1$ is nonempty, and so $r_1$ is big. Hence $m^+(vz)=1$ as required.

Suppose $k=5$. Then by hypothesis, $m(uv)=1$, $m(xy)=2$, and $r_2$ is small. We have $uv\in F_1,F_2$ and $xy \in F_3,F_4,F_5$.
Suppose that $uz \in Q_7$ and $m(uz) \geq 2$. Then $uz$ belongs to both $F_7$ and $F_6$. Further $vz\notin F_1,F_2,F_6,F_7$ and so by symmetry we can assume $vz \in F_5$. Consequently when $i\in I\setminus \{5\}$, we have $uz \in Q_i$, $n_i\geq 7$ and $e_3^i\l e_{n-1}^i$ belong only to $F_i$. Further, $m(uz)=2$. But then $F_i\cap Q_i\cap D_3$ is nonempty, contradicting the fact that $r_3$ is small. By the same argument if $m(vz)\geq 2$ then $vz \notin Q_7$.

Since $uvz$ is tough, by symmetry we may assume $m^+(uz) \geq 3$.  Thus $uz \notin Q_7$, and so $vz \in Q_7$ and $m(vz)=1$. Since $m(uz) \geq 2$,  $uz$ belongs to two of $F_3,F_4,F_5,F_6$; by symmetry say $uz\in F_5$. Thus for $i\in I\setminus \{5\}$, $vz \in Q_i$, $e_3^i\l e_{n_i-1}^i$ belong only to $F_i, F_6$. It follows that at least one of $F_i\cap Q_i \cap D_1$, $F_i\cap Q_i \cap D_2$ is nonempty, and if $i=6$ then both are nonempty. Thus $|D_1| + |D_2| \geq 7$, and since $r_2$ is small $|D_1| \geq 4$. It follows that $m^+(vz)=1$, as required.
This proves (1). \\

By (1) we may assume $m^+(vz) =1$.
Since $uvz$ is tough, (1) implies $m^+(uz) + m^+(uv) \geq 6$. Since $(G,m)$ does not contain Conf(\ref{c6}) by \ref{bigtri}, it follows that $m(uv)=2$, $m(uz)=2$ and $m(ux) \geq 3$. But $(G,m)$ does not contain Conf(\ref{c3}) by \ref{dense}, a contradiction.
This proves \ref{newsquare}.~\bbox

\begin{thm}\label{newc51}
No minimum $7$-counterexample contains Conf(\ref{c14}).
\end{thm}
\Proof
Let $(G,m)$ be a minimum $7$-counterexample, with a region $r$ bounded by a cycle $C_r=v_0\l v_4$. Denote the edge $v_iv_{i+1}$ by $f_i$ for $0\le i\le 4$ (taking indices modulo $5$) and suppose that $m^{+}(e_{0})\geq 2$, and that $m^{+}(f_{2}), m^{+}(f_{3})\geq 4$. Since $G$ has minimum degree at least three, $m(f_2) = m(f_3) = 3$.

Let $(G',m')$ be obtained by switching on the path $v_4\d v_0 \d v_1 \d v_2$; since $m(f_2), m(f_3)\geq 3$, $(G',m')$ contains a triangle $v_2v_3v_4$ with $m'(v_2v_3v_4) \geq 7$. Since $(G,m)$ is a $7$-target, $m(\delta_G(\{u,v,x\})) \geq 9$ and it follows that $m'(\delta_{G'}(\{u,v,x\})) \geq 7$. Since $m'(uv) + m'(ux) + m'(vx) \geq 7$, it follows that $m'(\delta(\{u,v,x\})) = 7$.
Hence by \ref{counterex}, $(G',m')$ is $7$-edge colourable. Let $F_1\l F_7$ be a $7$-edge colouring of $(G',m')$.
Let $k=m'(v_0v_1) + m'(v_2v_4) \geq 3$.
By \ref{cuts}
we may assume that $v_0v_1\in F_i$ for $1\le i\le m'(v_0v_1)$, and $v_2v_4\in F_k$.
Let $I = \{1\l 7\}\setminus\{k\}$ and for $i\in I$, let $Q_i$ be as in \ref{cuts}. 
Let $i\in I$, and let the edges of $Q_i$ in order be $e_1\l e_{n_i},e_1$, where $e_1 = v_0v_1$ and $e_2=v_2v_4$. Since $|F_i\cap Q_i|\ge 5$ and $F_i$ contains at most one of
$e_1,e_2$, it follows that $n_i\ge 6$.
For $1\le j\le 6$, $F_j$ contains one of $e_1,e_2$; and hence $e_3\l e_n\notin F_j$ for all $j\in \{1\l k\}$ with $j\ne i$. 
% It follows that $e_3\l,e_n$ only belong to $F_j$ for $j=i$ or $j>k$.
By the choice of the switchable path, $e_3 \in \{f_2,f_3\}$. By setting $i=7$, without loss of generality we may say $f_2\in Q_7$; it follows that $f_2$ does not belong to $F_1\l F_k$ and $k\leq 4$. Thus $f_2$ belongs to three of $F_{k+1}\l F_7$, say $f_2$ belongs to $F_5,F_6,F_7$. Further $f_3$ belongs to three of $F_1\l F_4$.
Let $r_2$ denote the second region for $f_2$ and let $D_2$ denote its set of doors.

It follows that $f_2\in Q_i$ for each $i\in I$. 
Suppose $k=4$. Then for each $i\in I$, the edges of $Q_i\setminus \{f_0,f_2\}$ belong only to $F_i$. Thus $F_i \cap Q_i \cap D_2$ is nonempty, contradicting the fact that $r_2$ is small.
Thus $k=3$, and so $m(f_1)=1$. Denote by $r_1$ the second region for $f_0$ and $D_1$ its set of doors. For each $i \in I$, $n_i \geq 7$ and the edges of $Q_i\setminus \{f_0,f_2\}$ belong only to $F_i, F_4$. Consequently at least one of $F_i \cap Q_i \cap D_1$, $F_i \cap Q_i \cap D_2$ is nonempty, and both are nonempty if $i=4$. Thus $|D_1| + |D_2| \geq 7$, but since $r_1$ is small, $|D_2|\geq 4$, a contradiction.
This proves \ref{newc51}.~\bbox

\begin{thm}\label{newc52}
No minimum $7$-counterexample contains Conf(\ref{c15}).
\end{thm}
\Proof

Let $(G,m)$ be a minimum $7$-counterexample, with a region $r$ bounded by a cycle $C_r=v_0\l v_4$. Denote the edge $v_iv_{i+1}$ by $f_i$ for $0\le i\le 4$ (taking indices modulo $5$) and suppose that $m^+(f_0)\geq 3$, and that $m^{+}(f_{2}), m^{+}(f_{3})\geq 3$. 
\\
\\
(1) {\em Suppose that either $f_0$ is $3$-heavy, or both $f_2,f_3$ are $3$-heavy. Then the path $v_4\d v_0 \d v_1 \d v_2$ is not switchable.}
\\
\\
Suppose the path $v_4\d v_0 \d v_1 \d v_2$ is switchable; let $(G',m')$ be obtained by switching on it and let $F_1 \l F_7$ be a $7$-edge colouring.
Let $k=m'(v_0v_1) + m'(v_2v_4) \geq 4$.
By \ref{cuts}
we may assume that $v_0v_1\in F_i$ for $1\le i\le m'(v_0v_1)$, and $v_2v_4\in F_k$.
Let $I = \{1\l 7\}\setminus\{k\}$ and for $i\in I$, let $Q_i$ be as in \ref{cuts}. 

Since $k\geq 4$ and $m(f_2), m(f_3) \geq 2$, we may assume without loss of generality that both $f_0, f_3$ belong to $F_1$.
Consequently, $f_2 \in Q_i$ for each $i\in I\setminus\{1\}$ and $f_2$ belongs to at least two of $F_{k+1}\l F_7$, say $f_2$ belongs to $F_6,F_7$, and so $k\leq 5$. 
Let $i\in I\setminus\{1\}$, and let the edges of $Q_i$ in order be $e_1\l e_n,e_1$, where $e_1 = v_0v_1, e_2=v_2v_4$ and $e_3 = f_2$. Since $|F_i\cap Q_i|\ge 5$ and $F_i$ contains at most one of
$e_1,e_2$, it follows that $n\ge 7$.
For $1\le j\le 6$, $F_j$ contains one of $e_1,e_2$; and hence $e_4\l e_n\notin F_j$ belong only to $F_i$, and possibly $F_7$. 

Denote by $r_1$, $r_2$ the second regions for $f_0, f_2$, respectively and denote by $D_1,D_2$ their respective sets of doors.
Suppose $k+m(f_2) =7$, and so $m(f_0) + m(f_2)\leq 5$. Then for each $i\in I\setminus \{1\}$, both $F_i\cap Q_i \cap D_1$, $F_i\cap Q_i \cap D_2$ are nonempty. It follows that both $r_1$ and $r_2$ are big, a contradiction.

Thus $k+m(f_2) \leq 6$, and so $k\leq 4$. For each $i\in I\setminus \{1\}$, at least one of $F_i\cap Q_i \cap D_1$, $F_i\cap Q_i \cap D_2$ is nonempty, and both are nonempty if $i=5$. Since at least one of $r_1, r_2$ is a triangle, one of $|D_1|, |D_2| \leq 2$, and so $k+m(f_2) \leq 6$.  $|D_1|+|D_2| \geq |I| =6$. But $k\geq 4$ and $m^+(f_2)\geq 3$ and so $r_1, r_2$ are both small, a contradiction.
This proves (1). \\

Now, suppose $(G,m)$ contains Conf(\ref{c15}), and so $f_0$ is $3$-heavy. By (1), the path $v_4\d v_0 \d v_1 \d v_2$ is not switchable, and $m(f_0) = 2$, and by symmetry we may assume $m(f_4)\geq 3$.
It follows that $m(f_2)\leq 2$, for otherwise we could relabel the vertices of $C_r$ to contradict (1). Further by (1) the path $v_1 \d v_2 \d v_3 \d v_4$ is not switchable.
Similarly $f_1$ is not $3$-heavy.
Since $v_1 \d v_2 \d v_3 \d v_4$ is not switchable, and $m(f_1), m(f_2) \leq 2$, it follows that $m(f_3) \geq 3$.
Further the $7$-target obtained by switching on $v_1 \d v_2 \d v_3 \d v_4$ contains Conf(\ref{c2}), and so by \ref{switchconf} it follows that $m(f_1) \geq 2$.
Now, the path $v_2 \d v_3 \d v_4 \d v_0$ is switchable; let $(G',m')$ be obtained by switching on it and let $F_1\l F_7$ be a $7$-edge-colouring.
Since $m'(v_3v_4) + m'(v_0v_2) = 5$, we may assume by \ref{cuts} that $v_3v_4$ belongs to $F_i$ for $1\leq i \leq 4$ and $v_0v_2 \in F_5$. Also by symmetry $v_2v_3$ and $v_4v_0$ both belong to $F_6$, and so $f_0, f_1$ do not belong to $F_6$.
Let $I = \{1\l 7\}\setminus \{5\}$ and for $i\in I$ let $Q_i$ be as in \ref{cuts}.
Let the edges of $Q_6$ in order be $e_1\l e_n,e_1$, where $e_1 = v_3v_4$ and $e_2=v_4v_0$. Since $|F_i\cap Q_6|\ge 5$ and $F_i$ contains at most one of
$e_1,e_2$, it follows that $n\ge 6$.
For $1\le j\le 6$, $F_j$ contains one of $e_1,e_2$; and hence $e_3\l e_n\notin F_j$ for all $j\in \{1\l k\}$ with $j\ne 6$. It follows that $e_3\l,e_n$ belong only to $F_6,F_7$.
By the choice of the switchable path, $e_3 \in \{f_0,f_1\}$, and so $m(e_3)\geq 2$. Hence $e_3$ belongs to both $F_6, F_7$, a contradiction.
This proves \ref{newc52}.~\bbox

\begin{thm}\label{sixgon}
No minimum $7$-counterexample contains Conf(\ref{c16}).
\end{thm}
\Proof
Let $(G,m)$ be a minimum $7$-counterexample, with a region $r$ bounded by a cycle $C_r=v_0\l v_5$. Denote the edge $v_iv_{i+1}$ by $f_i$ for $0\le i\le 5$ (taking indices modulo $6$) and suppose that $f_1,f_2,f_3,f_4,f_5$ are $3$-heavy with multiplicity at least two.
\\
\\
(1) {\em The path $v_0\d v_1\d v_2 \d v_3$ is not switchable.}
\\
\\
Suppose $v_0\d v_1\d v_2 \d v_3$ is switchable. Let $(G',m')$ be obtained by switching on it and let $F_1\l F_7$ be a $7$-edge-colouring of $(G',m')$.
Let $k=m'(v_1 v_2) + m'(v_0 v_3)\geq 4$. We may assume by \ref{cuts} that $v_1v_2\in F_i$ for $1\le i < k$ and $v_0v_3\in F_k$.
Let $I = \{1\l 7\}\setminus \{k\}$ and for $i\in I$, let $Q_i$ be as in \ref{cuts}. 

For $i\in I$, let the edges of $Q_i$ in order be $e_1^i\l e_{n_i}^i,e_1^i$, where $e_1^i = v_1v_2$ and $e_2^i = v_0v_3$. Since $|F_i\cap Q_i|\ge 5$ and $F_i$ contains at most one of
$e_1^i,e_2^i$, it follows that $n\ge 6$.
Let $i\in I$.
For $1\le j\le k$, $F_j$ contains one of $e_1^i,e_2^i$; and hence $e_3^i\l e_{n_i}^i\notin F_j$ for all $j\in \{1\l k\}$ with $j\ne i$. 
% Since $r$ is small, and by \ref{sixdoors} $(G,m)$ does not contain Conf(\ref{c11}), it follows that $k<7$. 
% Suppose that $k\geq 5$. Then $e_3^i\l e_n^i $ belong only to $F_i, F_6, F_7$. In particular $e_3^7$ is not $3$-heavy. 
By the choice of the switchable path $e_3^7 \in \{f_3,f_4,f_5\}$, and so $e_3^7$ is $3$-heavy; thus one of $e_3^7 e_4^7$ must belong to one of $F_1\l F_5$.

Thus $k=4$ and the second region for $v_1v_2$ is a triangle $v_1v_2x$. Choose $i\in \{5,6,7\}$ such that neither of $\{v_1x, v_2x\}$ is an edge of multiplicity one belonging to $F_i$. Now, $e_3^i\l e_{n_i}^i $ do not belong to $F_1\l F_4$. By the choice of the switchable path, $e_3^i$ is $3$-heavy, and so $e_{n_i}^i$ has multiplicity one and belongs only to $F_i$, a contradiction.
This proves (1). \\

Now $m(v_0v_1) \leq 2$, for otherwise the vertices of $C_r$ could be relabeled to contradict (1).
By (1), $v_0\d v_1\d v_2 \d v_3$ is not switchable. 
It follows that $m(v_1v_2)=2$ and the second region for $v_1v_2$ is a triangle and $m(v_2v_3)\geq 3$.
By symmetry, $m(v_5v_0)=2$, the second region for $v_5v_0$ is a triangle, and $m(v_4v_5) \geq 3$.
The $7$-target $(G,m)$ obtained by switching on $v_0\d v_1\d v_2 \d v_3$ contains Conf(\ref{c3}), so by \ref{switchconf} $(G,m)$ is smaller than $(G',m')$. It follows that $m(v_0v_1) + m(v_2v_3) \geq 5$. Similarly $m(v_0v_1) + m(v_4v_5) \geq 5$.

Since $m(v_2v_3)\geq 3$, the path $v_1\d v_2\d v_3\d v_4$ is switchable. Let $(G',m')$ be obtained by switching on it and let $F_1\l F_7$ be a $7$-edge-colouring.
Let $k=m'(v_2v_3) + m'(v_1v_4) \in \{5,6\}$.
We may assume by \ref{cuts} that $v_2v_3\in F_i$ for $1\le i < k$ and $v_1v_4\in F_k$. By symmetry we may assume $v_1v_2 \in F_{k+1}$.
Let $I = \{1\l 7\}\setminus \{k\}$ and for $i\in I$, let $Q_i$ be as in \ref{cuts}. 
Let the edges of $Q_7$ in order be $e_1\l e_n,e_1$, where $e_1 = v_2v_3$ and $e_2 = v_1v_4$. Since $|F_i\cap Q_i|\ge 5$ and $F_i$ contains at most one of $e_1,e_2$, it follows that $n\ge 6$.
For $1\le j\le k$, $F_j$ contains one of $e_1,e_2$; and hence $e_3\l e_n\notin F_j$ for all $j\in \{1\l k\}$ with $j\ne i$. 

Suppose $k=6$. Then $e_3\l e_n$ belong only to $F_7$, and so $e_3$ has multiplicity one. By the choice of the switchable path, $e_3 = f_0$.
But $f_0 \notin F_7$ since $f_1\in F_7$, a contradiction.
Thus $k=5$, and so $m(f_2)=3$ and $m(f_0)\geq 2$. Now $e_3\l e_n$ belong only to $F_6,F_7$, and so $e_3$ is not $3$-heavy. It follows from the choice of the switchable path that $e_3=f_0$. But $m(f_0)\geq 2$ and $f_0\notin F_6$ since $f_1 \in F_6$, a contradiction.
This proves \ref{sixgon}.~\bbox

This completes the proof of \ref{reduc} and hence of \ref{mainthm}.


\begin{thebibliography}{99}
\bibitem{appelhaken1} K.Appel and A.Haken, ``Every planar map is four colorable. Part I. Discharging'', {\em Illinois J. Math.} 21 (1977), 429--490.
\bibitem{appelhaken2} K.Appel, A.Haken and J.Koch, ``Every planar map is four colorable. Part II. Reducibility'', {\em Illinois J. Math.} 21 (1977), 491--567.
\bibitem{ces12} M. Chudnovsky, K. Edwards and P. Seymour, ``Edge-colouring eight-regular planar graphs'', manuscript (ArXiv 1209.1176), submitted.
\bibitem{dvorak} Z.Dvorak, K.Kawarabayashi and D.Kral, ``Packing six $T$-joins in plane graphs'', manuscript (2010arXiv1009.5912D)
\bibitem{katie} K.Edwards, {\em Optimization and Packings of $T$-joins and $T$-cuts}, M.Sc. Thesis, McGill University, 2011.
\bibitem{guenin} B.Guenin, ``Packing $T$-joins and edge-colouring in planar graphs'', {\em Mathematics of Operations Res.}, to appear.
\bibitem{rsst} N.Robertson, D.Sanders, P.Seymour and R.Thomas, 
``The four colour theorem'', {\em J. Combinatorial Theory, Ser. B}, 70 (1997), 2--44.
\bibitem{seymour} P. Seymour, {\em Matroids, Hypergraphs and the Max.-Flow 
Min.-Cut Theorem}, D.Phil. thesis, Oxford, 1975, page 34.
\end{thebibliography}
\end{document}